\def\gsim{\;\lower4pt\hbox{${\buildrel\displaystyle >\over\sim}$}\,}
\def\lsim{\;\lower4pt\hbox{${\buildrel\displaystyle <\over\sim}$}\,}

\def\avemag{$\langle{\bf B}\rangle$}
\def\FLASH{{\sc flash}}
\def\REMLIGHT{{\sc remlight}}
\def\PARAMESH{{\sc paramesh}}

\documentclass{aa}  
\usepackage{graphicx}
\usepackage{natbib}
\usepackage[english]{babel}
\usepackage{txfonts}
\newcommand\rs[1]{_\mathrm{#1}}
\begin{document}
   \title{Effects of non-uniform interstellar magnetic field on
          synchrotron X-ray and inverse-Compton $\gamma$-ray morphology
          of SNRs}

   \author{S. Orlando\inst{1},
           O. Petruk\inst{2,3},
           F. Bocchino\inst{1},
          \and
           M. Miceli\inst{4,1}
          }

   \offprints{S. Orlando,\\ e-mail: orlando@astropa.inaf.it}

   \institute{INAF - Osservatorio Astronomico di Palermo ``G.S.
              Vaiana'', Piazza del Parlamento 1, 90134 Palermo, Italy
         \and
              Institute for Applied Problems in Mechanics and
              Mathematics, Naukova St. 3-b Lviv 79060, Ukraine
         \and
              Astronomical Observatory, National University, Kyryla and
              Methodia St. 8 Lviv 79008, Ukraine
         \and 
              Dip. di Scienze Fisiche \& Astronomiche, Univ. di
              Palermo, Piazza del Parlamento 1, 90134 Palermo,
              Italy
             }

   \date{Received \quad\quad\quad ; accepted \quad\quad\quad }

   \authorrunning{S. Orlando et al.}
   \titlerunning{Effects of non-uniform ISMF on X-ray and
                 $\gamma$-ray morphology of SNRs}

 
  \abstract
   {Observations of SNRs in X-ray and $\gamma$-ray bands promise to
    contribute with important information in our understanding of the
    kinematics of charged particles and magnetic fields in the vicinity
    of strong non-relativistic shocks and, therefore, on the nature of
    galactic cosmic rays. The accurate analysis of SNRs images collected
    in different energy bands requires the support of theoretical
    modeling of synchrotron and inverse Compton emission from SNRs.}
   {We develop a numerical code (\REMLIGHT) to synthesize, from MHD
    simulations, the synchrotron radio, X-ray and inverse Compton
    $\gamma$-ray emission originating from SNRs expanding in non-uniform
    interstellar medium (ISM) and/or non-uniform interstellar magnetic
    field (ISMF). As a first application, the code is used to investigate
    the effects of non-uniform ISMF on the SNR morphology in the
    non-thermal X-ray and $\gamma$-ray bands.}
   {We perform 3D MHD simulations of a spherical SNR shock expanding
    through a magnetized ISM with a gradient of ambient magnetic field
    strength. The model includes an approximate treatment
    of upstream magnetic field amplification and the effect of shock
    modification due to back reaction of accelerated cosmic rays, assuming
    both effects to be isotropic. From the simulations, we synthesize
    the synchrotron radio, X-ray and inverse Compton $\gamma$-ray emission
    with the synthesis code \REMLIGHT, making different assumptions about
    the details of acceleration and injection of relativistic electrons.}
   {A gradient of the ambient magnetic field strength induces asymmetric
    morphologies in radio, hard X-ray and $\gamma$-ray bands independently
    from the model of electron injection if the gradient has a component
    perpendicular to the line-of-sight (LoS). The degree of asymmetry
    of the remnant morphology depends on the details of the electron
    injection and acceleration and is different in the radio, hard X-ray,
    and $\gamma$-ray bands. In general, the non-thermal X-ray morphology
    is the most sensitive to the gradient, showing the highest degree
    of asymmetry. The IC $\gamma$-ray emission is weakly sensitive to the
    non-uniform ISMF, the degree of asymmetry of the remnant morphology
    being the lowest in this band.}
   {}


   \maketitle

%
\section{Introduction}
\label{sec1}

It is largely accepted in the literature that the hard X-ray
emission detected in many young shell-type supernova remnants (SNRs)
is synchrotron emission from electrons accelerated to energies of tens
of TeV (e.g. \citealt{1995Natur.378..255K}) by means of the diffusive
shock acceleration process. In addition, inverse Compton (IC) collisions
of these high energy electrons with low energy photons from the ambient
radiation field (e.g. the cosmic microwave background; hereafter CMB)
are expected, thus leading to very-high-energy (VHE; $> 100$ GeV)
$\gamma$-ray emission too. In regions of high mass density, ions in
the shell are likely to have been accelerated to similar energies,
and $\gamma$-rays may be due to neutral pion decay from proton-proton
interactions. The nature of the TeV emission, therefore, stands on the
combination of X-ray synchrotron emitting electrons and very energetic
ions; it is not clear, at the present time, which one between the two
mainly contributes to the detected VHE $\gamma$-rays. The spectral
analysis of multi-wavelength data of several shell-type SNRs (Cas A,
\citealt{2001A&A...370..112A, 2007A&A...474..937A}; RX J1713.7-3946,
\citealt{2000A&A...354L..57M, 2002Natur.416..823E, 2007A&A...464..235A,
2008A&A...492..695B}; RX J0852.0-4622, \citealt{2005ApJ...619L.163K,
2005A&A...437L...7A, 2007ApJ...661..236A, 2006ApJ...652.1268E}; and RCW
86, \citealt{2008ICRC....2..585H}) allows both for leptonic and hadronic
origin of VHE $\gamma$-rays.

Recently, the study of SNRs as particle accelerators has received a
strong impulse thanks to the new $\gamma$-ray observations of SNRs with
the instruments of the High Energy Spectroscopic System (HESS), the
Major Atmospheric Gamma-ray Imaging Cherenkov (MAGIC) experiments and
the Fermi Gamma-ray Space Telescope.  The analysis of multi-wavelength
observations, from radio, to hard X-rays, to $\gamma$-rays, promises to
increase our understanding of the kinematics of charged particles and
magnetic fields in the vicinity of strong non-relativistic shocks and
of the possible role of SNRs for the origin of galactic cosmic rays
(CRs). In this context, a very important source of information could
be the distribution of surface brightness observed in SNRs in several
bands. For instance, the properties of the brightness distribution have
been crucial in our understanding of the acceleration and injection
of relativistic electrons through SNR shocks (e.g. the criterion of
\citealt{2004A&A...425..121R} versus the azimuthal profile comparison in
\citealt{2009MNRAS.393.1034P}). Recently, Petruk et al. (2010a, submitted
to MNRAS) have compared synthetic distributions of surface brightness
predicted for SN\,1006 in different bands with observations, deriving
important observational constraints on the modeling of SN\,1006. Also,
the correlations of brightness in radio, X-rays and $\gamma$-rays
claimed in RX J1713.7-3946 (\citealt{2009A&A...505..157A}) and some
others SNRs (e.g. \citealt{2006A&A...449..223A}) could be considered to
favor electrons as being responsible for VHE $\gamma$-rays in these SNRs.

The analysis of the brightness distributions observed in different
energy bands needs to be supported by an accurate comparison of the
observed distributions with those predicted by detailed MHD models. For
instance, a number of SNRs present a bilateral structure (BSNRs;
\citealt{1987A&A...183..118K, 1990ApJ...357..591F, 1998ApJ...493..781G})
and it is not clear how to ``translate'' this 2D information into a
3D morphology of the emission over the SNR shell. The two competing
cases traditionally invoked are either equatorial-belt or polar-caps and
they are related to the model of injection of relativistic electrons
(isotropic or quasi-perpendicular in the former case and quasi-parallel
in the latter case). Establishing the 3D morphology of SNRs, therefore,
may give some important hints on the acceleration theory.

The distribution of surface brightness of synchrotron emission in
SNRs expanding through a uniform interstellar medium (ISM) and
uniform interstellar magnetic field (ISMF) has been extensively
investigated, through numerical modeling, in both the radio
(\citealt{1990ApJ...357..591F}) and X-ray (\citealt{1998ApJ...493..375R,
2004AdSpR..33..461R}) bands; in particular, the dependence of the
brightness distributions on the efficiency of the acceleration process
has been explored, considering different injection models. First IC
$\gamma$-ray maps of SNRs in a uniform ISM and ISMF have been presented by
\cite{2009MNRAS.395.1467P} who investigated the properties of brightness
distributions in VHE $\gamma$-rays as compared to the distributions in
the radio band (see also Petruk et al. 2010b, submitted to MNRAS). Recently,
\cite{2009MNRAS.399..157P} proposed a method to predict IC $\gamma$-ray
images of SNRs starting from observed synchrotron radio maps and spatially
resolved X-ray spectral analysis (e.g. \citealt{2009A&A...501..239M})
of SNRs.

In a previous work (\citealt{2007A&A...470..927O}, hereafter Paper I),
we have investigated the origin of asymmetries in the radio morphology
of BSNRs through a model of a SNR expanding through either a non-uniform
ISM or a non-uniform ISMF. In this paper, we extend our analysis to the
non-thermal X-ray and IC $\gamma$-ray emission. In particular, we develop
a numerical code (\REMLIGHT) to synthesize the synchrotron radio, X-ray,
and IC $\gamma$-ray emission from 3D MHD simulations; then we couple the
synthesis code with the MHD model introduced in Paper I (extended
to include a simple treatment of upstream magnetic field amplification
and the effect of shock modification due to back reaction of accelerated
CRs) and investigate the effects of a non-uniform ISMF on the morphology
of the remnant in the hard X-ray and $\gamma$-ray bands. Though remnants
of Type Ia supernovae are expected to expand in an almost uniform ISMF,
here we show that even a very small gradient of the ISMF can influence
significantly the non-thermal remnant morphology.

In Sect.~\ref{sec2} we describe the MHD model and the numerical setup;
in Sect.~\ref{sec3} we describe the computation of synchrotron X-ray
and IC $\gamma$-ray emission, including model of relativistic electron
behavior; in Sect.~\ref{sec4} we discuss the results; and, finally,
in Sect.~\ref{sec5} we draw our conclusions.

\section{MHD modeling and numerical setup}
\label{sec2}

We adopt the MHD model introduced in Paper I, describing the propagation
of a SNR shock through a magnetized ambient medium. The shock propagation
is modeled by numerically solving the time-dependent ideal MHD equations
of mass, momentum, and energy conservation in a 3D Cartesian coordinate
system $(x,y,z)$ (see Paper I for details). The model does not include
consistently the effects on shock dynamics due to back-reaction of
accelerated CRs. However, we approach the effect of shock modification
by considering different values of the adiabatic index $\gamma$ which is
expected to drop from the value of an ideal monoatomic gas; in particular,
we consider here the cases of $\gamma=5/3$ (for an ideal monoatomic gas),
$\gamma = 4/3$ (for a gas dominated by relativistic particles), and
$\gamma = 1.1$ (for large energy drain from the shock region due to the
escape of high energy CRs). In addition, we account for upstream
magnetic field amplification due to back reaction of accelerated protons,
by amplifying the (pre-shock) ambient magnetic field in the neighborhoods
of the remnant. These effects, namely shock modification and magnetic
field amplification, might depend on the obliquity angle (i.e. the
angle between the external magnetic field and the normal to the shock);
for instance, they could follow the same dependence of the injection
efficiency (i.e. the fraction of accelerated electrons). However, at
the present time, the dependence of these effects on the obliquity angle
is not understood and, therefore, we consider here the simplest case of
isotropic shock modification and magnetic field amplification (i.e. no
additional obliquity-dependent magnetic-field amplification has been
assumed). On the other hand, such a choice makes easier our analysis
of non-thermal images which are already influenced by the obliquity
dependence of other processes (e.g. injection efficiency, magnetic
field compression, maximum energy of electrons; see Sect.~\ref{sec3}).
The simulations are performed using the \FLASH\ code (\citealt{for00}),
an adaptive mesh refinement multiphysics code for astrophysical plasmas.

As initial conditions, we adopt parameters appropriate to reproduce
the SNR SN\,1006 after 1000 yr of evolution: we assume an initial
spherical remnant with radius $r\rs{0snr} = 0.5$~pc, originating from a
progenitor star with mass of $1.4~M_{\rm sun}$, and propagating through
an unperturbed magneto-static medium. The initial total energy $E_0$ is
set to a value leading to a remnant radius $r\rs{snr} \approx 9$ pc at
$t=1000$ yr ($E_0$ ranges between $\approx 1.3$ and $1.8\times 10^{51}$
erg, depending on $\gamma$) and is partitioned so that most of the SN
energy is kinetic energy. The remnant expands through an homogeneous
isothermal medium with particle number density $n = 0.05$ cm$^{-3}$ and
temperature $T=10^4$ K. We consider three different configurations of
the unperturbed ambient magnetic field: 1) a uniform ambient magnetic
field (runs Unif-g1, Unif-g2, and Unif-g3); 2) a gradient of ambient
magnetic field strength perpendicular to the average magnetic field
(runs Grad-BZ-g1, Grad-BZ-g2, and Grad-BZ-g3); and 3) a gradient of
ambient magnetic field strength aligned with the average magnetic field
(runs Grad-BX-g1, Grad-BX-g2, Grad-BX-g3).

In the case of a uniform ISMF, we assume that the field is oriented
parallel to the $x$ axis. In the other two cases, the ambient magnetic
field is assumed to be dipolar\footnote{This idealized situation is
adopted here mainly to ensure magnetostaticity of the non-uniform
field.}. The dipole is oriented parallel to the $x$ axis and located
either on the $z$ axis ($x=y=0$) at $z=-100$ pc (Grad-BZ-g1, Grad-BZ-g2,
Grad-BZ-g3) or on the $x$ axis ($y=z=0$) at $x=-100$ pc (Grad-BX-g1,
Grad-BX-g2, Grad-BX-g3). In all the cases, we assume the magnetic
field strength to be $B_0 = 30~\mu$G at the center of the SN explosion
($x=y=z=0$), roughly an order of magnitude higher than the galactic
ISMF expected at the location of SN\,1006. This high value has been
chosen to mimic the effects of upstream magnetic field amplification
(see discussion above). In such a way, the post-shock ambient magnetic
field is expected to be of the order of a few hundred $\mu$G as deduced
from observations. In the configurations with non-uniform ISMF,
the field strength varies by a factor $\sim 6$ over 60 pc: either in
the direction perpendicular to the average ambient field {\avemag}
(Grad-BZ-g1, Grad-BZ-g2, Grad-BZ-g3), or parallel to {\avemag}
(Grad-BX-g1, Grad-BX-g2, Grad-BX-g3). We follow the expansion of the
remnant for 1000 yr. Table \ref{tab1} summarizes the physical parameters
characterizing the simulations considered here.

\begin{table}
\caption{Relevant initial parameters of the simulations.}
\label{tab1}
\begin{center}
\begin{tabular}{lcccc}
\hline
\hline
 & $\gamma$   & $E_0^a$         & magnetic field  & $(x,y,z)^b$ \\
 &            & [$10^{51}$ erg] & configuration   & pc  \\
\hline
Unif-g1     & 5/3      & 1.30   & uniform    & - \\
Unif-g2     & 4/3      & 1.54   & uniform    & - \\
Unif-g3     & 1.1      & 1.81   & uniform    & - \\
Grad-BZ-g1  & 5/3      & 1.30   & $z-$strat. & $(0,0,-100)$ \\
Grad-BZ-g2  & 4/3      & 1.54   & $z-$strat. & $(0,0,-100)$ \\
Grad-BZ-g3  & 1.1      & 1.81   & $z-$strat. & $(0,0,-100)$ \\
Grad-BX-g1  & 5/3      & 1.30   & $x-$strat. & $(-100,0,0)$ \\
Grad-BX-g2  & 4/3      & 1.54   & $x-$strat. & $(-100,0,0)$ \\
Grad-BX-g3  & 1.1      & 1.81   & $x-$strat. & $(-100,0,0)$ \\
\hline
\hline
\end{tabular}
\end{center}
$^a$\,Initial energy of the explosion. $^b$\,Coordinates
of the magnetic dipole moment.
\end{table}

The SN explosion is at the center $(x,y,z) = (0,0,0)$ of the computational
domain which extends between $-10$ and 10~pc in all directions. At the
coarsest resolution, the adaptive mesh algorithm used in the \FLASH\
code (\PARAMESH; \citealt{mom00}) uniformly covers the 3D computational
domain with a mesh of $8^3$ blocks, each with $8^3$ cells. We allow for
5 additional nested levels of refinement during the first 100 yr of
evolution with resolution increasing twice at each refinement level; then
the number of nested levels progressively decreases down to 2 at $t=1000$
yr as the remnant radius increases following the expansion of the
remnant through the magnetized medium. The refinement criterion adopted
(\citealt{loehner}) follows the changes in density and temperature. This
grid configuration yields an effective resolution of $\approx 0.0098$
pc at the finest level during the first 100 yr of evolution (when the
radius of the remnant was $<$ 2 pc) and $\approx 0.078$ pc at the
end of the simulation, corresponding to an equivalent uniform mesh of
$2048^3$ and $256^3$ grid points, respectively. We assume zero-gradient
conditions at all boundaries.

The model does not include the radiative cooling, describing
only the free and adiabatic expansion phases of the remnant.
The transition time from adiabatic to radiative phase for a SNR is
(e.g. \citealt{1998ApJ...500..342B, petruk05})

\begin{equation}
t\rs{tr} = 2.84\times 10^4\;E\rs{51}^{4/17}\;n\rs{ism}^{-9/17}~\mbox{yr}~,
\label{trans_time}
\end{equation}

\noindent
where $E\rs{51}=E\rs{0}/(10^{51}\ \mbox{erg})$ and $n\rs{ism}$ is
the particle number density of the ISM. In our set of simulations,
$t\rs{tr} > 10^5$ yr which is much larger than the time covered by
our simulations. Our modeled SNRs therefore never reach the radiative
phase. On the other hand, here we aim at describing young SNRs (i.e.
those that are still in the adiabatic expansion phase) from which
non-thermal X-ray and $\gamma$-ray emission is commonly detected.

\section{Synchrotron X-ray and inverse-Compton $\gamma$-ray emission
(REMLIGHT)} \label{sec3}

From the model results, we synthesize synchrotron radio, X-ray,
and IC $\gamma$-ray emission, by generalizing the approach of
\cite{1998ApJ...493..375R} to cases of non-uniform ISM and/or non-uniform
ISMF. In Paper I, we have already discussed the synthesis of synchrotron
radio emission and we refer the reader to that paper for the details of
calculation. Here we discuss the synthesis of X-ray and IC $\gamma$-ray
emission as it is implemented in the synthesis code \REMLIGHT.

One could assume that the synchrotron X-ray or IC $\gamma$-ray radiation
is due to relativistic electrons distributed with an energy spectrum
$N(E) = KE^{-s} \exp(-E/E\rs{max})$ electrons cm$^{-3}$ erg s$^{-1}$
(e.g. \citealt{1998ApJ...492..219G}), where $E$ is the electron
energy, $N(E)$ is the number of electrons per unit volume with arbitrary
directions of motion and with energies in the interval $[E, E+dE]$, $K$ is
the normalization of the electron distribution, $s$ the power law
index, and $E\rs{max}$ the maximum energy of electrons accelerated by the
shock\footnote{Note that the distribution of electrons $N(E)$ in the case
of synchrotron radio emission is expressed as $N(E) = KE^{-s}$ electrons
cm$^{-3}$ erg s$^{-1}$ (see, for instance, Paper I).}. Nevertheless,
some observations suggest that the cut-off could be broader than pure
exponent (e.g. \citealt{2000ApJ...540..292E, 2001ApJ...563..191E,
2003A&A...400..567U, 2004ApJ...602..271L}). Therefore we assume that
the energy spectrum of electrons is given by

\begin{equation}
N(E) = KE^{-s}
\exp\left[ -{\left(\frac{E}{E\rs{max}}\right)^{\alpha}}\right]~,
\label{el-distr}
\end{equation}

\noindent
where $\alpha\leq 1$ is the parameter regulating the broadening of the
high-energy end of electron spectrum\footnote{Note however
that recent work by \cite{2010ApJ...721..886K} suggests that
$\alpha>1$.}.

The volume emissivity due to synchrotron or IC radiation can be expressed as

\begin{equation}
i(\varepsilon) = \int_{0}^{\infty} N(E)\, \Lambda(E, \varepsilon, B)~dE~,
\label{emissivity}
\end{equation}

\noindent
where $\Lambda(E, \varepsilon, B)$ is the radiation power of a single
electron with energy $E$, and $\varepsilon$ is the photon energy. The
emissivity, $i(\varepsilon)$, depends on the magnetic field strength,
$B$, only in the synchrotron emission process. We compute the surface
brightness of the SNR at a given energy $\varepsilon$, by integrating
the emissivity $i(\varepsilon)$ at each point along each LoS in a
raster scan (assuming that the source is optically thin).

In the case of synchrotron emissivity in the X-ray band,
$i\rs{X}(\varepsilon)$, the spectral distribution of radiation power of
a single electron with energy $E$ in the magnetic field $\vec{B}$ is

\begin{equation}
\Lambda\rs{X}(E, \varepsilon) = \frac{\sqrt{3} e^3\mu_{\phi} B}{m\rs{e}c^2}\;
F\left(\frac{\varepsilon}{\varepsilon\rs{c}}\right)~,
\end{equation}

\noindent
where $\varepsilon\rs{c}=h\nu\rs{c} = hc_1\mu_{\phi} B E^2$, $h$ is the
Plank constant, $\nu\rs{c}$ is the critical frequency, $\phi$ the angle
between the magnetic field and the LoS, $\mu_{\phi}$ is either $\mu_{\phi}
= \sin \phi$ for the case of ordered magnetic field or $\mu_{\phi}
= \langle \sin \phi \rangle = \pi/4$ for disordered magnetic field,
$c_1 = 3e/(4\pi m\rs{e}^3c^5)$, $e$ and $m\rs{e}$ are the charge and
mass of electron, respectively, $c$ is the speed of light. The special
function $F(w)$ can be approximated as (e.g. \citealt{1985rpa..book.....R,
1959IAUS....9..595W}):

\begin{equation}
F(w) = \left\{
\begin{array}{ll}
  \displaystyle 2.15\, w^{1/3} &~~~~ w < 0.01~, \\
\\
  \displaystyle \sqrt{\pi}\, w^{0.29} \exp(-w) &~~~~ 0.01 \leq w \leq 5~,\\
\\
  \displaystyle \sqrt{\pi/2}\, w^{1/2} \exp(-w) &~~~~ w > 5~.\\
\end{array}\right.
\end{equation}

\noindent
We found the above approximation quite accurate with discrepancies
$\lsim 4\%$ from the exact value. In addition, $\int_0^{\infty}
F d\varepsilon = 1.59$ while the exact value is $8\pi/9\sqrt{3} = 1.61$.

In the case of $\gamma$-ray emissivity due to IC process, $i\rs{IC}$,
the spectral distribution $\Lambda\rs{IC}(E, \varepsilon)$ of
radiation power of a single electron in a black-body photon field in
Eq. \ref{emissivity} is (see also \citealt{2009MNRAS.395.1467P})

\begin{equation}
\Lambda\rs{IC}(E, \varepsilon) =
\frac{2e^4\epsilon\rs{c}}{\pi\hbar^3c^2} \Gamma^{-2}
{\cal{I}}\rs{ic}(\eta\rs{c},\eta\rs{0})~,
\end{equation}

\noindent
where $\Gamma$ is the Lorenz factor of electron, $\epsilon\rs{c} =
kT\rs{CMBR}$, $T\rs{CMBR}$ is the temperature of the cosmic microwave
background radiation (CMBR) assumed to be $T\rs{CMBR} =2.75$ K,

\begin{equation}
\eta\rs{c} = \frac{\epsilon\rs{c}\varepsilon}{(m\rs{e}c^2)^2}~,~~~~~~~~
\eta\rs{0}=\frac{\varepsilon^2}{4\Gamma m\rs{e}c^2(\Gamma m\rs{e}
c^2-\varepsilon)}~,
\end{equation}

\noindent
and the special function ${\cal{I}}\rs{ic}(\eta\rs{c},\eta\rs{0})$ can be 
accurately approximated as (\citealt{2009A&A...499..643P})

\begin{eqnarray}
\lefteqn{\displaystyle
{\cal{I}}\rs{ic}(\eta\rs{c},\eta\rs{0}) \approx
\frac{\pi^2}{6}\eta\rs{c} \left\{\exp\left[
-\frac{5}{4}\left(\frac{\eta\rs{0}}{\eta\rs{c}}\right)^{1/2}\right]\right. }
\nonumber \\
 & & ~~~~~~~~~~~~~ \left.  + 2\eta\rs{0} \exp
\left[-\frac{5}{7}\left(\frac{\eta\rs{0}}{\eta\rs{c}}\right)^{0.7}\right]
\right\}
 \exp\left[-\frac{2}{3}\left(\frac{\eta\rs{0}}{\eta\rs{c}}\right)
\right]~.
\end{eqnarray}

\noindent
This approximation represents ${\cal{I}}\rs{ic}(\eta\rs{c},\eta\rs{0})$
in any regime, from Thomson to extreme Klein-Nishina. The
approximation is exact in the Thomson limit. It restores detailed
calculations with maximum error of 30\% in the range of parameters which
gives non-negligible contribution to emission.

\subsection{Maximum energy of electrons}
\label{max_energy}

We follow the approach of \citet{1998ApJ...493..375R} for the
description of time evolution and surface variation of $E\rs{max}$,
generalizing his approach to cases of non-uniform ISM and/or
non-uniform ISMF. \citet{1998ApJ...493..375R} considered three
alternatives for time and spatial dependence of $E\rs{max}$. Namely,
the maximum accelerated energy maybe determined: 1) by the electron
radiative losses (due to synchrotron and IC processes), 2) by the
limited time of acceleration (due to the finite age of the remnant)
and 3) by properties of micro-physics when the scattering of electrons
with $E > E\rs{max}$ becomes less efficient and the electrons freely
escape from the region of acceleration\footnote{For the escape case, it
is commonly assumed that MHD waves responsible for the scattering are
much weaker above some wavelength, $\lambda\rs{max}$, and $E\rs{max}$
is approximately the energy of particles with that gyroradius
(e.g. \citealt{1998ApJ...493..375R}).}. The maximum energy is given by

\begin{equation}
E\rs{max,\xi} \propto f\rs{E,\xi}(\Theta\rs{o}) \; V\rs{sh}^{q\rs{\xi}}
\; B\rs{o}^{\lambda\rs{\xi}}~,
\label{emax}
\end{equation}

\noindent
where $\Theta\rs{o}$ is the obliquity angle, $f\rs{E,\xi}(\Theta\rs{o})$
is a function describing smooth variations of $E\rs{max}$ versus
obliquity, $V\rs{sh}$ is the shock velocity, $B\rs{o}$ is the pre-shock
ISMF strength, and $\xi=1,2,3$ corresponds respectively to loss-limited,
time-limited and escape-limited models of $E\rs{max}$. The values
of $q$ and $\lambda$ are: $q\rs{1} = 1$, $q\rs{2} = q\rs{3} = 0$,
and $\lambda\rs{1} = -1/2$, $\lambda\rs{2} = \lambda\rs{3} = 1$
(\citealt{1998ApJ...493..375R}). Note that we assume $q\rs{2}
= 0$ because 1) $E\rs{max}$ rises quite slowly with
time when it is determined by the finite time of acceleration
(\citealt{1998ApJ...493..375R}), even in the nonuniform ISM, and 2) most
of emission rises from downstream regions close to the shock; therefore,
the variation of $E\rs{max}$ due to velocity variation is negligible in
the (thin) emitting region. From Eq. \ref{emax}, we express the surface
variation of $E\rs{max}$ as

\begin{equation}
E\rs{max,\xi} = E\rs{max,\xi,\parallel}\; f\rs{E,\xi}(\Theta\rs{o})\;
{\cal{V}}\rs{sh}^{q\rs{\xi}} \;{\cal{B}}\rs{o}^{\lambda\rs{\xi}}~,
\label{emax_dep}
\end{equation}

\noindent
where $E\rs{max,\xi,\parallel}$ is a free parameter,
representing the maximum energy in a point $p$ on the
SNR surface where the ISMF is parallel to the shock normal,
${\cal{V}}\rs{sh} = V\rs{sh}/V\rs{sh,\parallel}$, ${\cal{B}}\rs{o} =
B\rs{o}/B\rs{o,\parallel}$, $V\rs{sh,\parallel}$ and $B\rs{o,\parallel}$
are the shock velocity and pre-shock ISMF strength in the point $p$,
respectively.

A detailed theoretical framework providing the obliquity dependence of
$E\rs{max}$ was presented by \cite{1998ApJ...493..375R} and is based on
the prescription for diffusion by \cite{1987ApJ...313..842J}. However,
this theory is limited to the test-particle regime, assuming no magnetic
field amplification.  On the other hand, at the present time, a more
general theory describing the obliquity dependence of $E\rs{max}$ is
still lacking. For the sake of generality, we adopt here some
arbitrary smooth variations of $E\rs{max}$ versus obliquity with the goal
to see how different trends in the obliquity dependence of $E\rs{max}$
influence the visible morphology of SNRs. In fact, the remnant morphology
-- once we are not interested in the exact comparison with observations
-- is mainly affected by the contrast ${\cal C}\rs{max} = E\rs{max,\|}
/ E\rs{max,\perp}$ and not by the exact shape of the dependence on
obliquity, once the latter is assumed to be smooth.

Our strategy is to consider smooth variations of $E\rs{max}$ versus
obliquity that correspond to the loss-limited, time-limited and
escape-limited models of $E\rs{max}$ in the theoretical framework
of \cite{1998ApJ...493..375R}. Since, in general, the mechanism
limiting electron acceleration (i.e. loss-limited, time-limited and
escape-limited) may be different at different shock obliquity angles,
we calculate the value of $E\rs{max}$ appropriate for each limitation
mechanism at each point, by considering $E\rs{max} = \min[E\rs{max,1},
E\rs{max,2}, E\rs{max,3}]$ (where indexes $1,2,3$ correspond respectively
to loss-limited, time-limited and escape-limited models). This way to
compute $E\rs{max}$ is adopted in Sect.~\ref{s_unif}, \ref{gamma_eff},
and \ref{var_emaxp}
where we assume also to be in the Bohm limit (i.e. gyrofactor\footnote{The
``gyrofactor'' is defined as the ratio between the mean free path,
$\lambda\rs{\parallel}$, along the magnetic field and the gyroradius,
$r\rs{g}$ (see \citealt{1998ApJ...493..375R}). In general it is expected
that the mean free path can be no less than $r\rs{g}$, so that $\eta
\geq 1$; the equality corresponds to the Bohm limit, i.e. a level of
turbulence leading to wave amplitudes comparable to the stationary
magnetic field strength.} $\eta = 1$) in the test-particle regime. In
particular, in Sect.~\ref{s_unif}, we introduce a reference case
for which the adopted set of parameters (see Sect.~\ref{par_space})
leads to ${\cal C}\rs{max}>1$. Note that the adopted parameters make
this case suitable for comparison with young non-thermal SNRs as,
for instance, SN\,1006. In addition, for the sake of generality, in
Sect.~\ref{sect_emax} and \ref{orient}, we explore the effects on the
remnant morphology of various obliquity dependencies of $E\rs{max}$, by
considering also cases for which the contrast ${\cal C}\rs{max}$ is $<1$.

\subsection{Post-shock evolution of the electron distribution}
\label{ps_el_distr}

As in \citet{1998ApJ...493..375R}, we assume that relativistic electrons
are confined in the fluid elements which advect them from the region of
acceleration. Fluid element with Lagrangian coordinate $\vec{a}\equiv
\vec{R}(t\rs{i})$ was shocked at time $t\rs{i}$, where $R$ is the radius
of the shock. At that time, the electron distribution on the shock was

\begin{equation}
N(E\rs{i}, t\rs{i}) = K\rs{s}(a,t\rs{i})E\rs{i}^{-s}\exp
\left[-\left(\frac{E\rs{i}}{E\rs{max}(t\rs{i})}\right)^{\alpha}\right]~,
\end{equation}

\noindent
where $E\rs{i}$ is the electron energy at time $t\rs{i}$, $K\rs{s}$
is the normalization of the electron distribution immediately after the
shock (in the following, index ``s'' refers to the immediately post-shock
values), and $s$ is the power law index. At variance with Paper I, we are
interested here in synchrotron X-ray and IC $\gamma$-ray emission. In this
case, the evolution of the electron distribution has to account for energy
losses of electrons due to both adiabatic expansion and radiative losses
caused by synchrotron and IC processes. At time $t\rs{i}$, the energy of
the electron confined in the fluid element with Lagrangian coordinates
$\vec{a}\equiv \vec{R} (t\rs{i})$ was (cf. Eq. 26 in
\citealt{1998ApJ...493..375R})

\begin{equation}
E\rs{i} = \frac{E}{{\cal{E}}\rs{ad}{\cal{E}}\rs{rad}}
\end{equation}

\noindent
where $E$ is the electron energy at the present time $t$, $\cal{E}\rs{ad}$
is a term accounting for the energy losses of electrons due to adiabatic
expansion

\begin{equation}
{\cal{E}}\rs{ad}(a,t) =
\left(\frac{\rho(a,t)}{\rho\rs{s}(t\rs{i})}\right)^{1/3} =
\left(\frac{\rho(a,t)}{\rho\rs{s}(t)}\right)^{1/3}
\left(\frac{\rho\rs{o}(R)}{\rho\rs{o}(a)}\right)^{1/3}
\end{equation}

\noindent
$\rho$ is the mass density (in the following, index ``o'' refers to
the pre-shock values), ${\cal{E}}\rs{rad}$ is a term accounting for the
radiative losses of electrons

\begin{equation}
{\cal{E}}\rs{rad}(E, a,t) = 1-{\cal{I}}(a,t)\frac{E}{E\rs{f,\parallel}}
\label{e_rad}
\end{equation}

\noindent
$E\rs{f,\parallel}= 637/(B\rs{eff,s,\parallel}^2 \,t)$ erg is the
fiducial energy at parallel shock (\citealt{1998ApJ...493..375R}),
$B\rs{eff,s,\parallel}^2 = B\rs{\parallel}^2+B\rs{CMB}^2$ is an
``effective'' magnetic field at parallel shock accounting energy losses
due to IC scatterings on the photons of CMB, $B\rs{CMB}=3.27$ $\mu$G is
the magnetic field strength with energy density equal to that in the
CMB, $t$ is the time, and ${\cal{I}}(a,t)$ is an integral independent
of $E$ which is calculated with the approach described in Appendix
\ref{app1}. The electron energy losses in a given fluid element are
mainly due to radiative losses if $E\rs{f} < E\rs{max}$ and to adiabatic
expansion if $E\rs{f} \gsim E\rs{max}$.

At time $t\rs{i}$, the shock was able to accelerate an electron
confined in the fluid element with Lagrangian coordinate $a$ to
$E\rs{max}(t\rs{i})$. From Eq. \ref{emax}, we derive that

\begin{equation}
\frac{E\rs{max}(t\rs{i})}{E\rs{max}(t)} =
\frac{f\rs{E}(\Theta\rs{o}(t\rs{i}))}{f\rs{E}(\Theta\rs{o}(t))}
\left(\frac{V\rs{sh}(t\rs{i})}{V\rs{sh}(t)}\right)^{q}
\left(\frac{B\rs{o}(a)}{B\rs{o}(R)}\right)^{\lambda} \equiv
{\cal{F}}(a,R)~.
\label{effe}
\end{equation}

\noindent
The ratio, $V\rs{sh}(t\rs{i})/V\rs{sh}(t)$, may be expressed through
pressure, $P$, and density, $\rho$ (\citealt{1999A&A...344..295H})

\begin{equation}
\frac{V\rs{sh}(t\rs{i})}{V\rs{sh}(t)} =
\left(\frac{P(a,t)}{P\rs{s}(t)}\right)^{1/2}
\left(\frac{\rho\rs{o}(a)}{\rho\rs{o}(R)}\right)^{(\gamma-1)/2}
\left(\frac{\rho(a,t)}{\rho\rs{s}(t)}\right)^{-\gamma/2}~.
\end{equation}

The conservation law for the number of particles per unit volume per
unit energy interval

\begin{equation}
N(E,a,t)=N(E\rs{i},a,t\rs{i}){a^2\,da\,dE\rs{i}\over \sigma
r^2\,dr\,dE},
\end{equation}

\noindent
where $\sigma=(\gamma+1)/(\gamma-1)$ is the shock compression ratio and
$r$ is the Eulerian coordinate, together with the continuity equation
$\rho_{\rm o}(a)a^2da=\rho(a,t)r^2dr$ and the derivative

\begin{equation}
\frac{dE\rs{i}}{dE} = \frac{1}{{\cal{E}}\rs{ad}{\cal{E}}^2\rs{rad}}~,
\end{equation}

\noindent
implies that downstream

\begin{eqnarray}
\lefteqn{N(E,a,t) = K(a,t)E^{-s}{\cal{E}}\rs{rad}^{s-2}}
\nonumber \\
 & & ~~~~~~~~~~~~~~~
\times 
\exp\left[-\left(\frac{E}{E\rs{max}(t,\Theta\rs{o})\,{\cal{F}}(a,R)\,
\cal{E}\rs{ad}\cal{E}\rs{rad}}
\right)^{\alpha}\right]~,
\label{deriv_distrib}
\end{eqnarray}

\noindent
with $K(a,t) = K\rs{s}(t\rs{i})\,{\cal{E}}\rs{ad}^{s+2}$,
$E\rs{max}(t,\Theta\rs{o})$ is given by Eq.~\ref{emax_dep}, and
taking into account Eq. \ref{effe}. Assuming that $K\rs{s}\propto
\rho\rs{s}V\rs{sh}(t)^{-b}$, i.e. it varies with the shock velocity
$V\rs{sh}(t)$ and, in case of non-uniform ISM, with the immediately
post-shock value of mass density, $\rho\rs{s}$, the downstream variation
of $K(a,t)$ is described by the relation (see Paper I)

\begin{eqnarray}
\lefteqn{\frac{K(a,t)}{K\rs{s}(R,t)}=
\frac{f\rs{\varsigma}(\Theta\rs{o}(t\rs{i}))}{f\rs{\varsigma}(\Theta\rs{o}(t))}
\;\left(\frac{P(a,t)}{P\rs{s}(R,t)}\right)^{-b/2}}
\nonumber \\
 & & ~~~~~~~
\times \left(\frac{\rho\rs{o}(a)}{\rho\rs{o}(R)}\right)^{-b(\gamma-1)/2-(s-1)/3}
 \left(\frac{\rho(a,t)}{\rho\rs{s}(R,t)}\right)^{b\gamma/2+(s+2)/3}~,
\label{kappa}
\end{eqnarray}

\noindent
where $f\rs{\varsigma}(\Theta\rs{o})$ is the obliquity dependence of
the injection efficiency $\varsigma$ (the fraction of accelerated
electrons). Again, in the lack of theoretical dependence
of the injection efficiency on obliquity in case of efficient
acceleration, we consider a number of smooth functions for
$f\rs{\varsigma}(\Theta\rs{o})$, exploring either increasing or
decreasing injection. In fact, such an approach, which considers
different contrasts $f\rs{\varsigma\|}/f\rs{\varsigma\bot}$ either $>1$
or $<1$ is realized in our previous paper \cite{2009MNRAS.395.1467P};
there, it was shown how the change in the injection contrast influences
non-thermal images of SNRs in uniform ISM and uniform ISMF. In the
present paper, for simplicity and in order to see the effects from
nonuniformity of ISMF, we limit our considerations to three contrasts of
injection. Namely, following \citet{1998ApJ...493..375R}, we consider
the following models: quasi-parallel ($f\rs{\varsigma}(\Theta\rs{o})
= \cos^2 \Theta\rs{s}$), isotropic ($f\rs{\varsigma}(\Theta\rs{o}) =
1$), and quasi-perpendicular ($f\rs{\varsigma}(\Theta\rs{o}) = \sin^2
\Theta\rs{s}$), where $\Theta\rs{s}$ is related to $\Theta\rs{o}$
through the expression $\cos \Theta\rs{s} = \sigma\rs{B}^{-1} \cos
\Theta\rs{o}$. The first injection model leads to a three-dimensional
polar-caps structure of the remnant, whereas the latter two produce a
three-dimensional equatorial-belt structure of the remnant.

Radiative losses of electrons $\dot E \propto E^2$ are mostly
effective in modification of the distribution $N(E,a,t)$ around
$E\sim E\rs{max}$ (\citealt{1998ApJ...493..375R}). This may 
be noted in Eq.~\ref{deriv_distrib}. The variation of the energy
distribution $N(E,a,t)$ of electrons with energy $E\ll E\rs{max}$
(in this case also $E\ll E\rs{f}$, leading to $\cal{E}\rs{rad}
\rightarrow$ 1), i.e. electrons with negligible radiative losses,
is given by $N(E,a,t)/N\rs{s}(E,R,t) = K(a,t)/K\rs{s}(R,t)$, where
$N\rs{s}(E,R,t)$ is the energy distribution of electrons immediately
after the shock.  This expression does not depend on energy $E$ and,
in fact, we have used this expression in Paper I for investigation
of properties of surface brightness distribution of SNR emitting radio
frequencies. In contrast, the modification of the distribution $N(E,a,t)$
due to effective electron radiation is given by the two last multipliers
in Eq.~\ref{deriv_distrib}. The radiative losses of electrons therefore
are important for the surface brightness distribution of SNR in X-ray
and $\gamma$-rays.


\section{Synchrotron and IC images of SNRs expanding through
non-uniform ISMF}
\label{sec4}

\begin{figure}[!t]
  \centering
  \includegraphics[width=8.5cm]{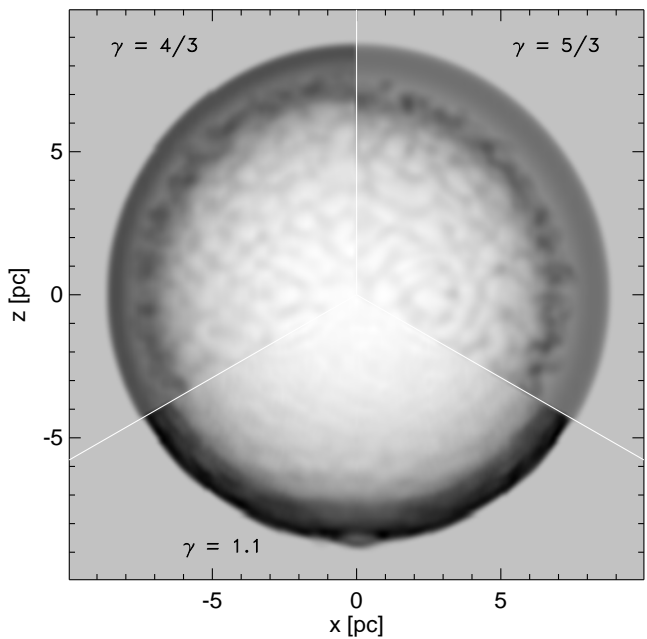}
  \caption{3D rendering of the mass density at $t=1000$ yr for a
           remnant expanding through a uniform ISMF and for three cases
           of $\gamma = 5/3,\,4/3,\,1.1$ (runs Unif-g1, Unif-g2, and
           Unif-g3, see Table~\ref{tab1}).}
  \label{fig2}
\end{figure}

The evolution of the remnant expanding through the non-uniform ISMF
has been described in Paper I where the reader is referred to for more
details. Figure~\ref{fig2} shows the 3D rendering of the mass density
at $t=1000$ yr in the three cases of $\gamma$ considered for uniform
ISMF. The main effect of $\gamma$ on the shock dynamics is to change its
compression ratio and the distance of the contact discontinuity from the
blast wave position; no dependence on the obliquity angle is present,
$\gamma$ being uniform in each simulation. The value of $\gamma$ is
expected therefore to influence the absolute values of emission in the
radio, X-ray and $\gamma$-ray bands but not the large scale morphology
of the remnant to which this paper is focused on. In the following,
we first discuss the effects of non-uniform ISMF on the synchrotron
and IC emission adopting, as reference, the case with $\gamma = 5/3$,
allowing the direct comparison of our results with those available in
the literature; then in Sect.~\ref{gamma_eff}, we discuss the effect of
$\gamma$ on the morphology of the non-thermal emission.

In all the synthetic images presented below, we introduce the
procedure of magnetic field disordering (with randomly oriented
magnetic field vector in each point) downstream of the shock
(see Paper I), according to observations showing a low degree of
polarization (10-15\%; e.g. Tycho, \citealt{1991AJ....101.2151D},
SN\,1006, \citealt{1993AJ....106..272R}). Note that the obliquity
angle $\Theta\rs{s}$ is derived from $\Theta\rs{o}$ through the
conversion formula $\cos \Theta\rs{s} = \sigma\rs{B}^{-1}
\cos \Theta\rs{o}$ given in Sect.~\ref{ps_el_distr} and, therefore,
does not take into account the magnetic field disordering; as discussed
by \cite{1990ApJ...357..591F}, this corresponds to the assumption that
the disordering process takes place over a longer time-scale than the
electron injection which occurs in the close proximity of the shock.

\begin{figure}[!t]
  \centering
  \includegraphics[width=7.5cm]{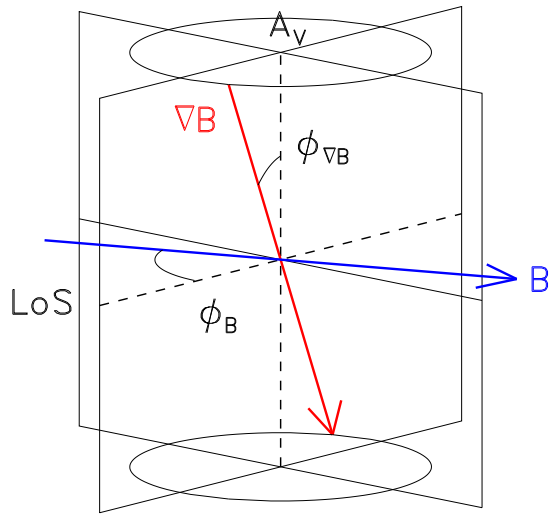}
  \caption{Relevant angles describing the orientation of the ISMF
           and of the gradient of ISMF strength with
           respect to the observer: $\phi\rs{B}$ is the angle between the
           (average) unperturbed ISMF and the LoS, and $\phi\rs{\nabla
           B}$ is the angle between the gradient of the ISMF
           strength and the vertical line passing through the center of
           the remnant $A\rs{V}$.}
  \label{fig3}
\end{figure}

In all the simulations, we assume the (average) unperturbed ISMF {\avemag}
oriented along the $x$ axis. In the two magnetic field configurations
explored in this paper, the gradient of ISMF strength is either normal
(runs Grad-BZ-g1, Grad-BZ-g2, Grad-BZ-g3; $\nabla |$\avemag$|$ along $z$)
or aligned (runs Grad-BX-g1, Grad-BX-g2, Grad-BX-g3; $\nabla |$\avemag$|$
along $x$) to {\avemag}. Since we analyze the remnant morphology as it
would be observed from different points of view, we define two angles
to describe the orientation of {\avemag} and $\nabla |$\avemag$|$
in the space (see Fig.~\ref{fig3}): $\phi\rs{B}$ is the angle between
{\avemag} and the LoS, and $\phi\rs{\nabla B}$ is the angle between
$\nabla |$\avemag$|$ and the normal to the ISMF in the plane of the sky
(axis $A\rs{v}$ in Fig.~\ref{fig3}). The first angle is the aspect angle
commonly used in the literature. The definition of the second angle allows
us to explore the remnant morphology for various aspect angles and for
fixed $\phi\rs{\nabla B}$, $\nabla |$\avemag$|$ lying on a cone with
angle $\phi\rs{\nabla B}$ (see Fig.~\ref{fig3}).  In cases in which the
gradient $\nabla B$ is aligned with the average ISMF (runs Grad-BX-g1,
Grad-BX-g2, Grad-BX-g3) $\phi\rs{\nabla B}=90^0$ by definition. In Grad-BZ
models, the angle between {\avemag} and $\nabla |$\avemag$|$ is always
$90^0$. In the following, the images are calculated for various values of
the angles defined above and with a resolution of $256\times 256$ pixels.

\subsection{Parameter space}
\label{par_space}

The prescriptions for the electron energy distribution at any point inside
the remnant and for the synthesis of synchrotron and IC emission discussed
in Sect.~\ref{sec3} are characterized by several parameters regulating
the energy spectrum of relativistic electrons, the injection efficiency,
the time and spatial dependence of $E\rs{max}$, etc.. In the following,
we limit the model parameter space through some assumptions that allow
us to fix some of the parameters.

In particular, we assume that the power law index in
Eq.~\ref{el-distr} is $s=2$, as suggested by many observations of BSNRs
(e.g. for SN\,1006; \citealt{2009A&A...501..239M}). In the test-particle
regime, the index should be related to the shock compression ratio through
$s=(2+\sigma)/(\sigma-1)$ with $\sigma=(\gamma+1)/(\gamma-1)$. In case
of efficient shock acceleration, the electron energy distribution is
curved. The value of $s$ that should be used in an approximation like
Eq.~\ref{el-distr} is rather a mean radio-to-X-ray spectral index whose
value is around 2 (e.g. \citealt{2008ApJ...683..773A}), independent
of the local slopes of the electron spectrum. As for the curvature
of the spectrum around $E\rs{max}$, we assume that $\alpha = 0.5$
in Eq.~\ref{el-distr}. In fact, this could be the case in SN\,1006
and G347.3-0.5 where a number of models suggest $\alpha \approx 0.5$
(\citealt{2000ApJ...540..292E, 2001ApJ...563..191E, 2003A&A...400..567U,
2004ApJ...602..271L}).

The maximum energy at parallel shock $E\rs{max, \parallel}$ is a free
parameter in Eq.~\ref{emax_dep} that we assume to be $E\rs{max,\parallel}
= 26$ TeV in most of our calculations. This parameter has
to be compared with the fiducial energy at parallel shock defined
as $E\rs{f,\parallel}= 637/(B\rs{eff,s,\parallel}^2 \,t)$ erg
(\citealt{1998ApJ...493..375R}); for the cases considered here,
we have $E\rs{f,\parallel}= 14$ TeV in models with uniform ISMF,
$E\rs{f,\parallel}= 12$ TeV in Grad-BZ models and $E\rs{f,\parallel}=
2$ TeV in Grad-BX models\footnote{Note that $E\rs{f,\parallel}$
depends on the magnetic field strength at parallel shock which is
different in the three configurations of unperturbed ISMF explored
here due to the magnetic field gradient (in all the cases, we fix
the magnetic field strength at the center of the SN explosion).}. In
all these cases, therefore, $E\rs{f} < E\rs{max}$ for a significant
portion of the remnant and the electron energy losses are mainly due to
radiative losses (see discussion in Sect.~\ref{ps_el_distr}). In
Sect.~\ref{var_emaxp}, we investigate the dependence of the non-thermal
emission on $E\rs{max,\parallel}$, by exploring cases for which $E\rs{max}
< E\rs{f}$ and the electron energy losses are mainly due to adiabatic
expansion.

The parameter $b$ in Eq.~\ref{kappa} is a constant and determines how
the injection efficiency depends on the shock properties; we assumed that
$K\rs{s}\propto \rho\rs{s}V\rs{sh}(t)^{-b}$ (see Sect.~\ref{ps_el_distr}
and Paper I). On theoretical grounds $b$ might be expected to be negative,
reflecting an expectation that injection efficiency may behave in a way
similar to acceleration efficiency: stronger shocks might inject particles
more effectively. \cite{1998ApJ...493..375R} considered three empirical
alternatives for $b$ as a free parameter, namely, $b=0,-1,-2$. In
particular, $b=-2$ is commonly assumed in many areas of astrophysics such
as gamma-ray bursts and prompt radio and X-ray emission from SNe. However
\cite{2010A&A...509A..34B} have shown that models preferring a constant
fraction of the shock energy to be transferred into CRs (i.e. $b=-2$)
are rejected by statistical analysis of two SNR samples. In addition,
Petruk et al. (2010a, submitted to MNRAS) compared their model results
with experimental data of the remnant SN\,1006 and found that $b$ has
a value between 0 and $-1$. Petruk et al. (2010b, submitted to MNRAS)
showed that the smaller $b$, the thicker the radial profiles of the
surface brightness in all bands; an effect mostly prominent in radio
band. Since no effect on the pattern of asymmetries induced
by a non-uniform ISMF is expected (see Paper I), we assume $b=0$ in all
our calculations, this being the most neutral case.

The synthetic images are expected to depend on the remnant age
(\citealt{1998ApJ...493..375R}). To reduce further the number of
model parameters, we focus here on remnant 1 kyr old, as in the case of
SN\,1006. Finally, radio, X-ray and $\gamma$-ray images are synthesized at
1 GHz, 3 keV, and 1 TeV, respectively. It is worth to emphasize that all
the above parameters are not expected to influence the pattern
of the asymmetries induced by a non-uniform ISMF on which the present
paper is focused (see, also, Sect.~\ref{var_emaxp} for a discussion on
the influence of $E\rs{max,\parallel}$ on the remnant morphology).

\subsection{Asymmetries in the remnant morphology: the reference case}
\label{s_unif}

\begin{figure}[!t]
  \centering
  \includegraphics[width=8.4cm]{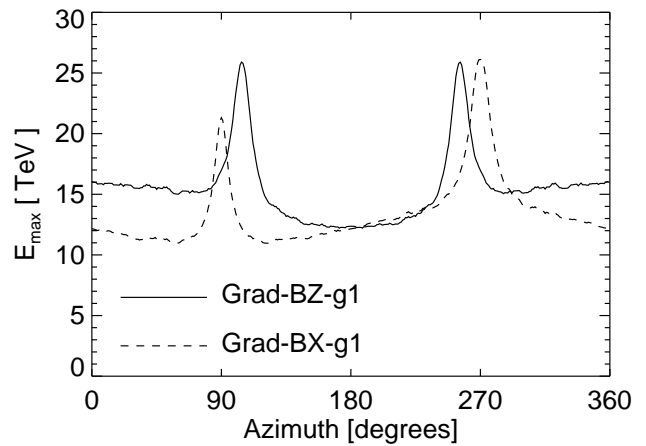}
  \caption{Azimuthal profiles of the maximum energy $E\rs{max}$
           computed in run Grad-BZ-g1 (solid line) and Grad-BX-g1
           (dashed line) when the aspect angle is $\phi\rs{B} = 90^{0}$
           and the gradient of magnetic field strength lies in the plane
           of the sky ($\phi\rs{\nabla B} = 0^0$ in run Grad-BZ-g1 and
           $\phi\rs{\nabla B} = 90^0$ in run Grad-BX-g1). The shock is
           parallel around $90^0$ and $270^0$ and perpendicular around
           $180^0$ and $360^0$. In both models, the contrast ${\cal
           C}\rs{max} > 1$.}
  \label{fig1}
\end{figure}

In Paper I, we analyzed the asymmetries induced by a non-uniform
ISMF in the radio morphology of the remnant. In particular, we found
there that asymmetric BSNRs are produced if a gradient of the ambient
magnetic field strength $\nabla B$ is not aligned with the LoS. In
this section we extend our analysis to non-thermal X-rays and IC
$\gamma$-rays. To this end, we synthesize the synchrotron and IC emission,
considering each of the three cases of variation of electron injection
efficiency with shock obliquity (quasi-perpendicular, isotropic, and
quasi-parallel particle injection). Also we assume the adiabatic index
to be $\gamma=5/3$; the effects of lower $\gamma$ values on the remnant
morphology are explored in Sect.~\ref{gamma_eff}. Here the maximum energy of
electrons is calculated at each point as $E\rs{max} = \min[E\rs{max,1},
E\rs{max,2}, E\rs{max,3}]$ (where indexes $1,2,3$ correspond respectively
to loss-limited, time-limited and escape-limited models; see discussion
in Sect.~\ref{max_energy}). For the set of parameters chosen for our
simulations, it turns out that the loss-limited model is dominant at all
obliquity angles, thus simplifying the analysis of non-thermal images in
this section. Figure~\ref{fig1} shows $E\rs{max}$ versus the azimuthal
angle (the azimuth is measured counterclockwise from the "north" of the
remnant) for runs Grad-BZ-g1 and Grad-BX-g1. In both cases, $E\rs{max}$
is characterized by two maxima where the ISMF is parallel to the shock
normal (around $90^0$ and $270^0$); thus the contrast of $E\rs{max}$
is ${\cal C}\rs{max} > 1$. The strength of the unperturbed ISMF is the
largest at $180^0$ ($90^0$) and the lowest at $360^0$ ($270^0$) in run
Grad-BZ-g1 (Grad-BX-g1) due to the magnetic field gradient. The latter
determines the asymmetries in the azimuthal profile of $E\rs{max}$:
in run Grad-BZ-g1, the two maxima are converging on the side where
the field is the most intense, the gradient being perpendicular to
the average magnetic field; in run Grad-BX-g1, the two maxima have
different intensities (with the largest where the magnetic field is the
lowest\footnote{In the loss-limited model $E\rs{max} \propto B^{-1/2}$,
see Sect.~\ref{max_energy}.}), the gradient being parallel to the average
magnetic field.

\begin{figure}[!t]
  \centering
  \includegraphics[width=8.8cm]{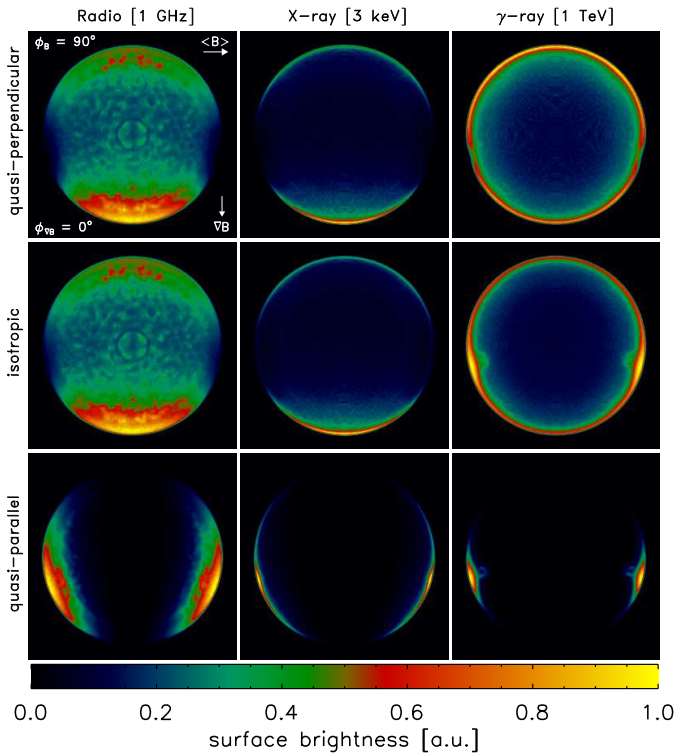}
  \caption{Maps of synchrotron radio (left), X-ray (center), and IC
           $\gamma$-ray (right) surface brightness (normalized to the
           maximum of each map) at $t=1$ kyr synthesized from run
           Grad-BZ-g1, assuming randomized internal magnetic field. The
           relevant angles are $\phi\rs{B} = 90^{0}$ and $\phi\rs{\nabla
           B} = 0^0$. The figure shows the quasi-perpendicular (top),
           isotropic (middle), and quasi-parallel (bottom) particle
           injection models.  The adiabatic index is $\gamma=5/3$. The
           average ambient magnetic field is along the horizontal axis;
           the gradient of magnetic field strength is along the vertical
           axis.}
  \label{fig4}
\end{figure}
\begin{figure}[!t]
  \centering
  \includegraphics[width=8.8cm]{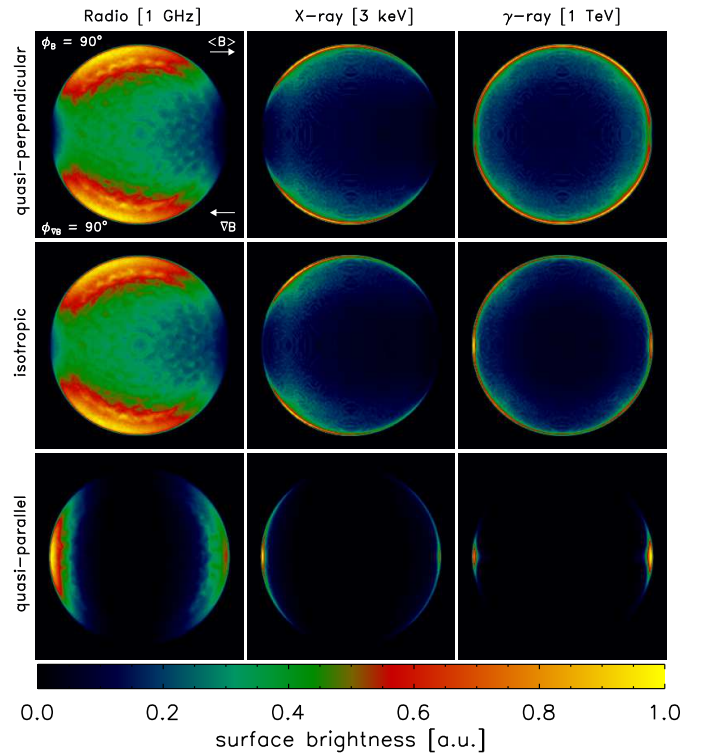}
  \caption{As in Fig.~\ref{fig4} for run Grad-BX-g1. Both the average
           ambient magnetic field and the gradient of magnetic field
           strength are along the horizontal axis. The relevant angles
           are $\phi\rs{B} = 90^{0}$ and $\phi\rs{\nabla B} = 90^0$.}
  \label{fig5}
\end{figure}

As an example, Figs.~\ref{fig4} and \ref{fig5} show the maps of
synchrotron radio, X-ray, and IC $\gamma$-ray surface brightness at
$t=1$ kyr, in each of the three injection models (quasi-perpendicular,
isotropic, and quasi-parallel). The aspect angle is $\phi\rs{B} = 90^{0}$
in all images, i.e. the ambient magnetic field is perpendicular to the
LoS; the angle $\phi\rs{\nabla B}$ is $0^0$ for run Grad-BZ-g1 and $90^0$
for Grad-BX-g1.

The main factors affecting the azimuthal variations of
surface brightness are the variations of: injection efficiency
$f\rs{\varsigma}(\Theta\rs{o})$ and magnetic field $B\rs{s}(\Theta\rs{o})$
in the radio band; $f\rs{\varsigma}(\Theta\rs{o})$,
$B\rs{s}(\Theta\rs{o})$ and maximum energy $E\rs{max}(\Theta\rs{o})$
in the X-ray band; $f\rs{\varsigma}(\Theta\rs{o})$ and
$E\rs{max}(\Theta\rs{o})$ in the IC $\gamma$-ray band. Therefore, the
morphology of the remnant in the three bands can differ considerably
in appearance. In the radio and in the X-ray band, the remnant shows
two lobes located at perpendicular shocks in the quasi-perpendicular
and isotropic models (i.e. where the magnetic field is larger), and
at parallel shocks in the quasi-parallel model (i.e. where emitting
electrons reside). The lobes are much thinner in X-rays than in radio
because of the large radiative losses at the highest energies that
make the X-ray emission dominated by radii closest to the shock. In
the $\gamma$-ray band, the remnant morphology changes significantly
in the three injection models: it is almost ring-like (with two faint
minima at parallel shocks) when the injection is quasi-perpendicular;
the morphology shows two lobes located at parallel shocks when the
injection is isotropic, at variance with the lobes in radio and X-rays
that are located at perpendicular-shocks (i.e. bright $\gamma$-ray
lobes correspond to dark radio and X-ray areas); the morphology is
characterized by two narrow bright lobes almost superimposed to those
in radio and X-rays when the injection is quasi-parallel. A ring-like
$\gamma$-ray morphology is compatible with those found by HESS in the
SNRs RX J1713.7-3946 (\citealt{2006A&A...449..223A}) and RX J0852.0-4622
(Vela Jr.; \citealt{2007ApJ...661..236A}) where $\gamma$-rays are detected
virtually throughout the whole remnant and the emission is found to
resemble a shell structure. On the other hand, the bipolar $\gamma$-ray
morphology of SN\,1006 revealed by HESS (\citealt{2010A&A...516A..62A}),
with the bright lobes strongly correlated with non-thermal X-rays, may
be easily reproduced in the polar-caps scenario (quasi-parallel injection).

The effects of the non-uniform ISMF on the remnant morphology in the
X-ray band are similar to those discussed in Paper I for the radio band:
remnants with two non-thermal X-ray lobes of different brightness (upper
left panel in Fig.~\ref{fig4} and lower left panel in Fig.~\ref{fig5})
are produced if a gradient of ambient magnetic field strength is
perpendicular to the lobes; remnants with converging similar non-thermal
X-ray lobes (lower left panel in Fig.~\ref{fig4} and upper right panel
in Fig.~\ref{fig5}) are produced if the gradient runs between the two
lobes. Analogous asymmetries are found in the $\gamma$-ray morphology
of the remnant although the degree of asymmetry is less evident. Note
however that, in the case of isotropic injection, the $\gamma$-ray lobes
are converging on one side when radio and X-ray lobes are characterized by
different brightness (see Fig.~\ref{fig4}). This is the consequence of the
``limb-inverse'' property in $\gamma$-rays\footnote{The ``limb-inverse''
property in $\gamma$-rays is determined for isotropic injection
because the magnetic field affects the downstream distribution of IC
$\gamma$-ray emitting electrons which is steeper where the magnetic
field is stronger. The reader is referred to \cite{2009MNRAS.395.1467P}
for more details.} (\citealt{2009MNRAS.395.1467P}). In general, this
property is valid not only in the case of isotropic injection; this
type of injection is just the more prominent case. In fact, the critical
quantities determining the ``limb-inverse'' property are the contrasts
between electron injection, ISMF, and model of $E\rs{max}$. For instance,
in the case of uniform ISMF, the azimuthal contrast in IC $\gamma$-ray
brightness is roughly

\begin{figure*}[!t]
  \sidecaption
  \includegraphics[width=12.cm]{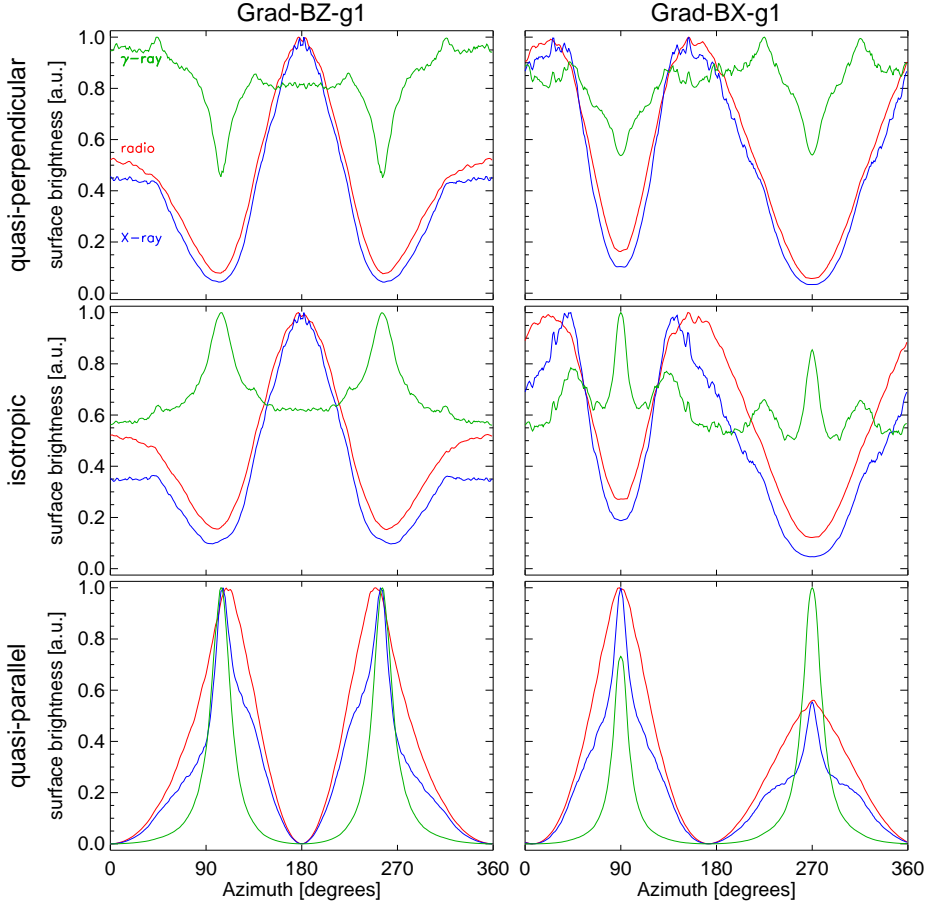}
  \caption{Azimuthal profiles of the synchrotron radio (red), X-ray
           (blue), and IC $\gamma$-ray (green) surface brightness
           synthesized from runs Grad-BZ-g1 (left; the relevant angles
           are $\phi\rs{B} = 90^{0}$ and $\phi\rs{\nabla B} = 0^0$) and
           Grad-BX-g1 (right; $\phi\rs{B} = 90^{0}$ and $\phi\rs{\nabla
           B} = 90^0$), assuming quasi-perpendicular (top), isotropic
           (middle), and quasi-parallel (bottom) injection models.
           The adiabatic index is $\gamma=5/3$. The azimuth
           is measured counterclockwise from the north (see
           Figs.~\ref{fig4} and \ref{fig5}). The corresponding azimuthal
           profiles of $E\rs{max}$ used to derive the curves of this
           figure are in Fig.~\ref{fig1}.}
  \label{fig6}
\end{figure*}

\begin{equation}
\begin{array}{rl}
\displaystyle 
\frac{S\rs{\parallel}}{S\rs{\perp}} & \displaystyle
\propto \frac{\rm injection\rs{\parallel}}{\rm injection\rs{\perp} }
\exp\left[-E\rs{m}\left(\frac{1}{E\rs{max,\parallel}} -
\frac{1}{E\rs{max,\perp}}\right)\right] = \\ \\
  &  \displaystyle
 = \frac{\rm injection\rs{\parallel}}{\rm injection\rs{\perp}}
\exp\left[-\frac{E\rs{m}}{E\rs{max,\parallel}}
\left(1-\frac{E\rs{max,\parallel}}{E\rs{max,\perp}}\right)\right]\\
\end{array}
\end{equation}

\noindent
where $E\rs{m}$ is the electron energy which gives the maximum
contribution to IC emission at a considered frequency and subscripts
refer to positions along the limb where the ambient magnetic
field is either parallel ($\parallel$) or perpendicular ($\perp$)
to the shock normal. Even in the case of quasi-parallel injection
($\rm injection\rs{\parallel}/injection\rs{\perp} > 1$), the contrast
$S\rs{\parallel}/S\rs{\perp}$ depends on the contrast of $E\rs{max}$:
the ratio ${\cal C}\rs{max} = E\rs{max,\parallel}/E\rs{max,\perp}$
may lead to an exponential term either $>1$ or $<1$, leading to
$S\rs{\parallel}/S\rs{\perp}$ either $>1$ or $<1$.

Another interesting feature characterizing the IC $\gamma$-ray morphology
of the remnant is the inversion of the asymmetry when the two lobes
have different brightness (i.e. a gradient of magnetic field strength
is perpendicular to the lobes). This feature is evident in the upper
panels of Fig.~\ref{fig4} and in the lower panels of Fig.~\ref{fig5}: the
brightest $\gamma$-ray lobe is located where both the radio and the X-ray
lobes are fainter. As discussed in detail below in Sect.~\ref{sect_emax},
this is due to the fact that, in the synthetic images presented in this
section, $E\rs{max}$ depends inversely on the pre-shock ambient magnetic
field strength (see Eq.~\ref{emax_dep} and Fig.~\ref{fig1})
and its contrast is ${\cal C}\rs{max} > 1$.

Figure \ref{fig6} shows the azimuthal profiles of the synchrotron
radio, X-ray, and IC $\gamma$-ray surface brightness synthesized
from runs Grad-BZ-g1 and Grad-BX-g1 for the three injection models
when the relevant angles are $\phi\rs{B}=90^0$ and $\phi\rs{\nabla
B}=0^0$ for run Grad-BZ-g1 and $\phi\rs{\nabla B}=90^0$ for
Grad-BX-g1. In the quasi-parallel scenario, the non-thermal
lobes are rather narrow azimuthally. Note the ``limb-inverse''
property in $\gamma$-rays for isotropic injection as discussed by
\cite{2009MNRAS.395.1467P}. Note also the ``asymmetry-inverse''
property in $\gamma$-rays when the two lobes have different brightness. In
general we find that the degree of asymmetry (whatever the pattern of
asymmetry -- either different brightness or convergence of the lobes --
is) induced by $\nabla B$ in the remnant morphology is different in the
three bands (see Sect.~\ref{orient} for a discussion on the degree of
asymmetry of the remnant in the different bands); in particular, the IC
$\gamma$-ray emission appears to be the less sensitive to the gradient.

Useful parameters to quantify the degree of asymmetry of the remnant are
those defined in Paper I: the azimuthal intensity ratio $R\rs{max}\geq
1$, i.e. the ratio of the maxima of intensity of the two lobes as derived
from the azimuthal intensity profiles (a measure of different brightness
of the lobes; $R\rs{max} > 1$ in case of asymmetry), and the azimuthal
distance $\theta\rs{D}$, i.e. the distance in deg of the two maxima (a
measure of the convergence of the lobes; $\theta\rs{D} < 180^0$ in case
of asymmetry). For instance, in the case of quasi-parallel injection
in Fig.~\ref{fig6} (lower panels), we find that the azimuthal distance
$\theta\rs{D}$ ranges from $148^0$ in $\gamma$-rays and X-rays
to $134^0$ in radio for run Grad-BZ-g1, and the azimuthal intensity
ratio $R\rs{max}$ ranges from 1.4 in $\gamma$-rays to 1.8 in radio and
X-rays for run Grad-BX-g1.

\subsection{Dependence on the adiabatic index}
\label{gamma_eff}

\begin{figure*}[!t]
  \centering
  \includegraphics[width=18.cm]{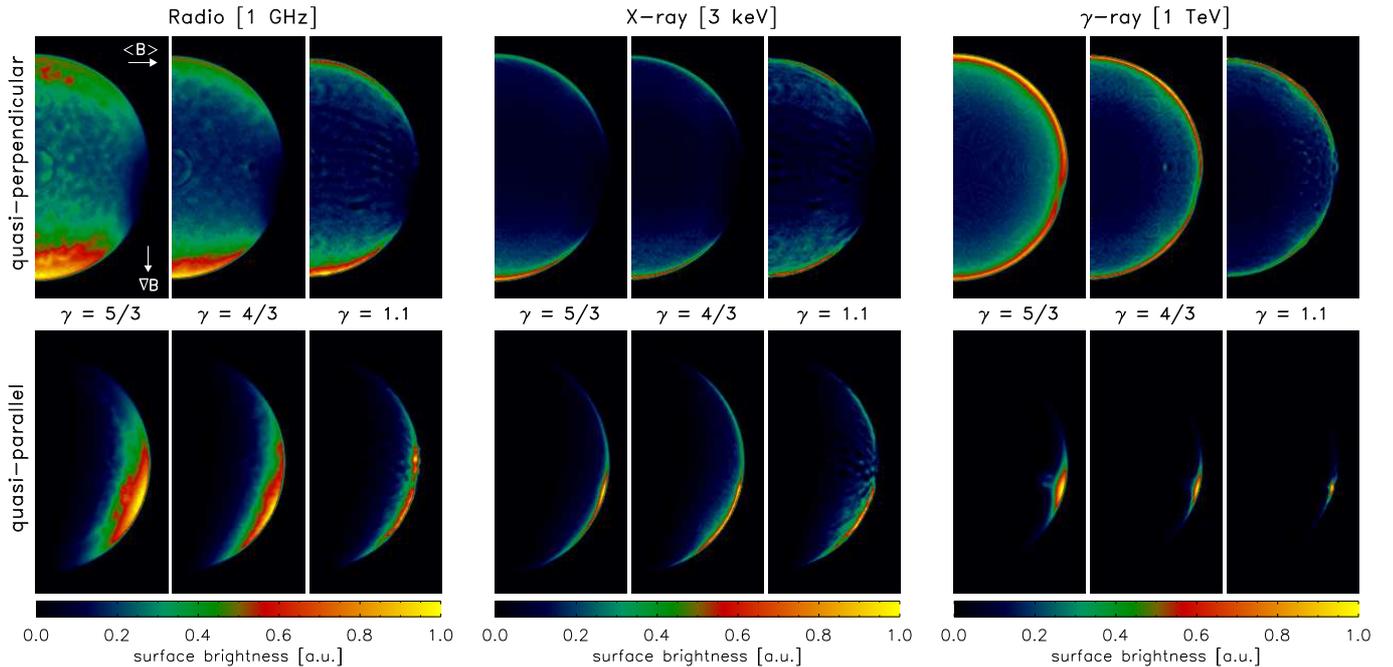}
  \caption{As in Fig. \ref{fig4} for runs Grad-BZ-g1 ($\gamma = 5/3$),
           Grad-BZ-g2 ($\gamma = 4/3$), and Grad-BZ-g3 ($\gamma =
           1.1$). The figure shows synchrotron radio (left panels),
           X-ray (center panels) and IC $\gamma$-ray (right panels),
           assuming either quasi-perpendicular (upper panels) or
           quasi-parallel (lower panels) injection models.
           Each panel shows only one half of the remnant which
           is symmetric with respect to the vertical axis.}
  \label{fig7}
\end{figure*}

Petruk et al. (2010b, submitted to MNRAS) analyzed the effect of $\gamma$
on non-thermal images of SNR expanding through homogeneous ISM and uniform
ISMF. They showed that reducing the value of $\gamma$, the synchrotron
brightness of the remnant is modified by increased radiative losses of
emitting electrons, due to increased compression of $\vec{B}$, which
results in thinner radial profiles of brightness. Figure \ref{fig7} shows
maps of synchrotron radio, X-ray, and IC $\gamma$-ray emission for the
case of ISMF characterized by a gradient of field strength perpendicular
to the average magnetic field and different values of the adiabatic index
$\gamma$ (runs Grad-BZ-g1, Grad-BZ-g2, and Grad-BZ-g3). As expected, the
index $\gamma$ determines both the shock compression ratio $\sigma$ and
the distance of the contact discontinuity from the blast wave position
$D\rs{cd}$ (see also Fig~\ref{fig2} in the case of uniform ISMF): the
smaller $\gamma$, the larger $\sigma$ (and the larger the radiative
losses of emitting electrons) and the shorter $D\rs{cd}$. As shown in
the figure, the main effect of smaller $\gamma$ is to make thinner the
lobes emitting synchrotron emission in the three bands. In particular,
in the extreme case of $\gamma = 1.1$, the lobes are so thin that they
are largely perturbed by the hydrodynamic instabilities forming at
the contact discontinuity, the typical size of the instabilities being
comparable with $D\rs{cd}$. The adiabatic index slightly influences also
the azimuthal thickness of the lobes, especially in the quasi-parallel
case: the smaller $\gamma$, the narrower this thickness. Nevertheless,
the adiabatic index does not change significantly neither the degree
nor the pattern of asymmetry of the remnant morphology caused by the
gradient of magnetic field strength.

\subsection{Dependence on the maximum energy}
\label{sect_emax}

\begin{figure}[!t]
  \centering
  \includegraphics[width=8.2cm]{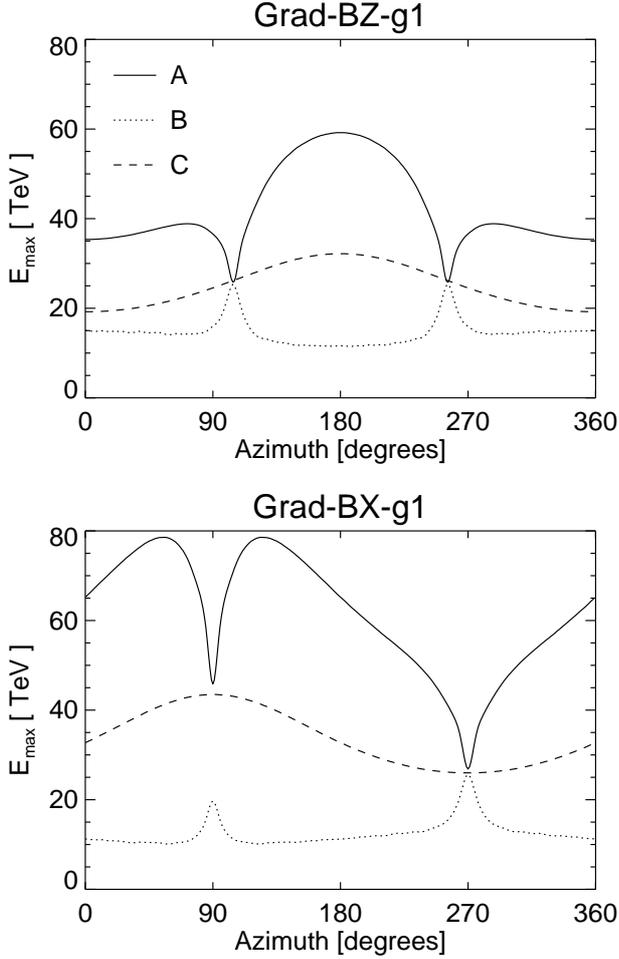}
  \caption{As in Fig.~\ref{fig1} but for three different
           models of $E\rs{max}$ characterized by smooth variations with
           the obliquity angle. The azimuth is measured counterclockwise
           from the "north" of the remnant. The shock is parallel
           around $90^0$ and $270^0$ and perpendicular around $180^0$ and
           $360^0$. In model A (solid line), the contrast of $E\rs{max}$
           is ${\cal C}\rs{max} < 1$, in model B (dotted) ${\cal
           C}\rs{max} > 1$, and in model C (dashed) ${\cal C}\rs{max}$
           is $< 1$ on the side of the remnant with the strongest ISMF
           strength and $> 1$ on the side with the lowest field strength
           (see text).}
  \label{fig8bis}
\end{figure}

\begin{figure}[!t]
  \centering
  \includegraphics[width=8.cm]{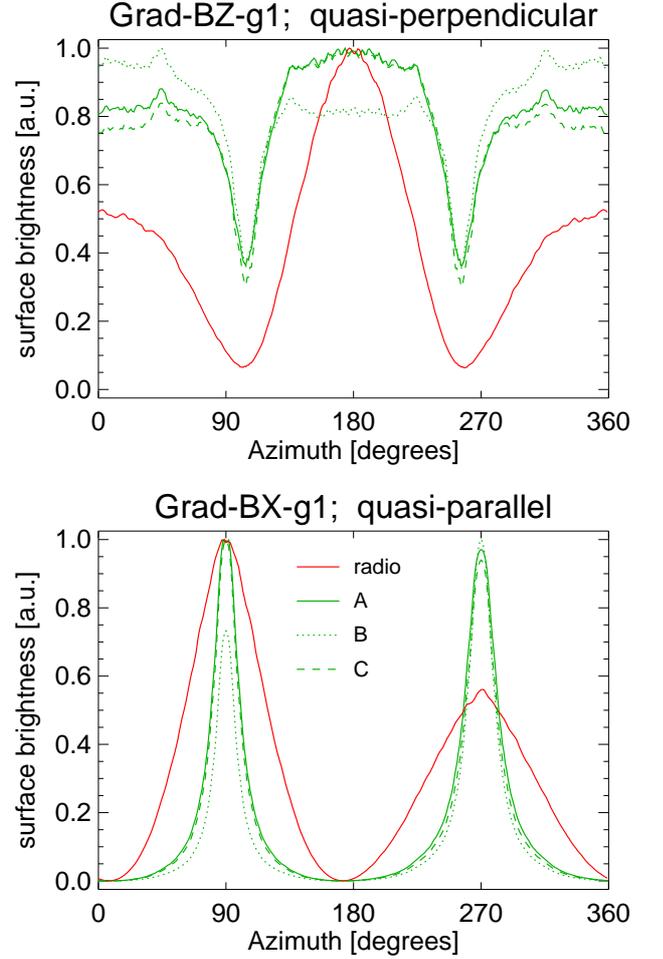}
  \caption{Azimuthal profiles of the synchrotron radio (red lines)
           and IC $\gamma$-ray surface brightness (green) synthesized
           from run Grad-BZ-g1 with quasi-perpendicular injection (top),
           and from run Grad-BX-g1 with quasi-parallel injection (bottom);
           the aspect angle is $\phi\rs{B} = 90^0$. The models
           of $E\rs{max}$ adopted to synthesize the non-thermal emission
           are those shown in Fig.~\ref{fig8bis}: model A (solid line),
           B (dotted) and C (dashed).}
  \label{fig8}
\end{figure}

In Sect.~\ref{s_unif}, we have presented the remnant morphology for
a reference case for which $E\rs{max}$ calculated at each point as
$E\rs{max} = \min[E\rs{max,1}, E\rs{max,2}, E\rs{max,3}]$ (where
indexes $1,2,3$ correspond respectively to loss-limited, time-limited
and escape-limited models; see discussion in Sect.~\ref{max_energy}) is
characterized by a contrast ${\cal C}\rs{max} > 1$. Here we generalize
our study by considering some arbitrary smooth variations of $E\rs{max}$
versus obliquity with the goal to see how different trends in the
obliquity dependence of $E\rs{max}$ influence the visible morphology
of SNRs. That is, we do not use here the particular prescriptions
for $E\rs{max}$ from Sect. ~\ref{max_energy}, but simply assume that
the acceleration physics is able to operate to produce an $E\rs{max}$
of prescribed properties. In fact, the model of $E\rs{max}$ and its
obliquity dependence may affect both the degree and the pattern of
asymmetry of the remnant morphology. A critical point is if $E\rs{max}$
depends directly or inversely on the magnetic field strength. Moreover
the remnant morphology in the various bands can be characterized by
different features if the contrast ${\cal C}\rs{max}$ is either $>1$
or $<1$. As an example, Fig.~\ref{fig8bis} shows the azimuthal profiles
of three arbitrary models of $E\rs{max}$ characterized by different
dependencies on the obliquity angle: in model A, $E\rs{max}\propto B$
and its contrast is ${\cal C}\rs{max} < 1$ (solid line); in model B,
$E\rs{max}\propto B^{-1/2}$ and ${\cal C}\rs{max} > 1$ (dotted); in model
C, $E\rs{max}\propto B$ and ${\cal C}\rs{max}$ is $< 1$ on the side of
the remnant with the strongest ISMF strength and $> 1$ on the side with
the lowest field strength. Note that model B coincides with the model
of $E\rs{max}$ computed for the reference case in Sect.~\ref{s_unif}
(see Fig.~\ref{fig1}). The asymmetries on the profiles of $E\rs{max}$
are introduced by the gradient of ambient magnetic field, as discussed
in Sect.~\ref{s_unif}.

From the models of $E\rs{max}$ shown in Fig.~\ref{fig8bis}, we synthesized
maps of synchrotron radio, X-ray, and IC $\gamma$-ray emission. We find
that the remnant morphology is very sensitive to the model of $E\rs{max}$
when a gradient of magnetic field strength is perpendicular to the lobes
and the latter are characterized by different brightness. The asymmetry
between the two lobes can be reduced in the X-ray band or even inverted
in the IC $\gamma$-ray band when $E\rs{max}$ depends inversely on the
pre-shock ambient magnetic field strength, namely in the case of model
B in Fig.~\ref{fig8bis}. In particular, this model of $E\rs{max}$ leads
to the ``asymmetry-inverse'' property in $\gamma$-rays already discussed
in Sect.~\ref{s_unif} for our reference case (see upper right panel in
Fig.~\ref{fig4} and lower right panel in \ref{fig5}).

Fig.~\ref{fig8} shows the azimuthal profiles of the IC $\gamma$-ray
surface brightness synthesized from runs Grad-BZ-g1 and Grad-BX-g1 when
the lobes have different brightness, for the three models of $E\rs{max}$
reported in Fig.~\ref{fig8bis}. In the case of the IC surface brightness,
the asymmetry-inverse property is evident when $E\rs{max}$ is the
largest where the magnetic field strength is the lowest (see model B in
Fig.~\ref{fig8bis}). This is due to the fact that the
IC emissivity $i(\epsilon)$ weakly depends on $\vec{B}$. In the case
of the non-thermal X-ray surface brightness, the inverse dependence
of $E\rs{max}$ on $\vec{B}$ partially contrast the dependence of the
non-thermal X-ray $i(\epsilon)$ on $\vec{B}$, reducing the degree of
asymmetry between the lobes. It is worth to note that, if $E\rs{max}$
is high enough in regions with weak magnetic field, than the inversion
of asymmetry may be present even in the X-ray band.

On the other hand, we also found that when the non-uniform ISMF leads to
non-thermal lobes converging on one side (i.e. when a gradient of ISMF
is running between the lobes) the model of $E\rs{max}$ does not affect
significantly the degree and the pattern of asymmetry of the remnant
morphology (see lower panel in Fig.~\ref{fig8}).

\subsection{Dependence on the orientation of ISMF gradient}
\label{orient}

\begin{figure}[!t]
  \centering
  \includegraphics[width=8.7cm]{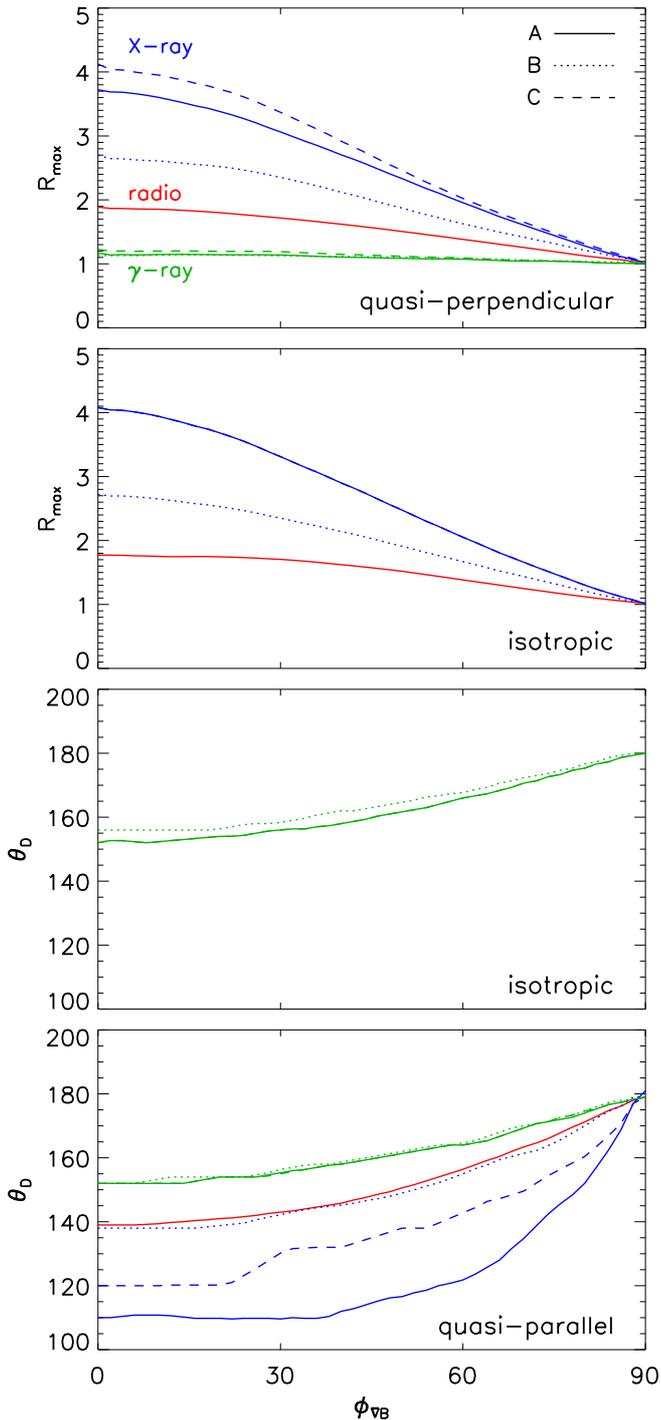}
  \caption{Azimuthal intensity ratio $R\rs{max}$ (i.e. the ratio of the
           maxima of intensity of the two lobes around the shell) and
           azimuthal distance $\theta\rs{D}$ (i.e. the distance in deg of
           the two maxima of intensity around the shell) vs. the angle
           between $\nabla B$ and the vertical line passing through
           the remnant center $\phi\rs{\nabla B}$, for an aspect angle
           $\phi\rs{B} = 90^0$, and for the three models of
           $E\rs{max}$ shown in Fig.~\ref{fig8bis}. The run is
           Grad-BZ-g1. For isotropic injection, curves for $R\rs{max}$
           in $\gamma$-rays and curves for $\theta\rs{D}$ in radio and
           X-rays are not shown, the values being $R\rs{max} = 1$ and
           $\theta\rs{D}=180^o$, respectively at all $\phi\rs{\nabla B}$.}
  \label{fig10}
\end{figure}

As expected, the degree of asymmetry of the remnant morphology depends on
the orientation of $\nabla B$ with respect to the plane of the sky. In the
case of run Grad-BZ-g1, Fig.~\ref{fig10} shows the azimuthal intensity
ratio $R\rs{max}$ and the azimuthal distance $\theta\rs{D}$ vs. the
angle $\phi\rs{\nabla B}$, for an aspect angle $\phi\rs{B} = 90^0$
and for different trends in the obliquity dependence of $E\rs{max}$
(exploring the contrasts ${\cal C}\rs{max}$ either $>1$ or
$<1$; see Fig.~\ref{fig8bis}). The asymmetries are the largest when
$\nabla B$ lies in the plane of the sky (i.e. $\phi\rs{\nabla B} =
0^0$), whereas no asymmetries are present when $\nabla B$ is along the
LoS (i.e. $\phi\rs{\nabla B} = 90^0$). In all the intermediate cases,
the degree of asymmetry is determined by the component of $\nabla B$
lying in the plane of the sky. Note that the remnant morphology shows
only one kind of asymmetry when the injection is quasi-perpendicular or
quasi-parallel and the aspect angle is $\phi\rs{B} = 90^0$. On the other
hand, the lobes have different brightness in radio and non-thermal X-rays
and are converging in IC $\gamma$-rays when the injection is isotropic
due to the "limb-inverse" property.

In general we find that the degree of asymmetry (whatever the pattern of
asymmetry -- either different brightness or convergence of the lobes --
is) induced by $\nabla B$ in the remnant morphology is different in the
three bands: the non-thermal X-ray (IC $\gamma$-ray) emission appears to
be the most (less) sensitive to the gradient. This happens because the
emissivity $i(\epsilon)$ depends directly on the magnetic field strength
(see Eq.~\ref{emissivity}) only in the synchrotron emission process
(no in the IC process). Consequently, the IC $\gamma$-ray emission
shows a weaker dependence on the $\nabla B$. In fact, IC brightness
depends on $\vec{B}$ indirectly, through radiative losses of electrons:
larger $\vec{B}$ induces decrease of the number of electrons emitting
IC $\gamma$-rays. Note that the sensitivity on $\nabla B$ depends also
on the energy of photons, and on the reduced fiducial energy $E\rs{f}$,
which is the measure of efficiency of the role of radiative losses in
modification of the downstream evolution of emitting electrons. Note also
that the degree of asymmetry of the remnant morphology can be
significantly reduced when the $E\rs{max}$ contrast is ${\cal C}\rs{max} >
1$ (e.g. model B in Fig.~\ref{fig8bis}). In fact, the asymmetry
reduction is mainly due to the dependency of $E\rs{max}$ on the magnetic
field strength. As discussed in Sect.~\ref{sect_emax}, the asymmetries in
the remnant morphology can be reduced or even inverted when $E\rs{max}$
depends inversely on the pre-shock ambient magnetic field strength
(which is the case in model B).

\begin{figure}[!t]
  \centering
  \includegraphics[width=8.8cm]{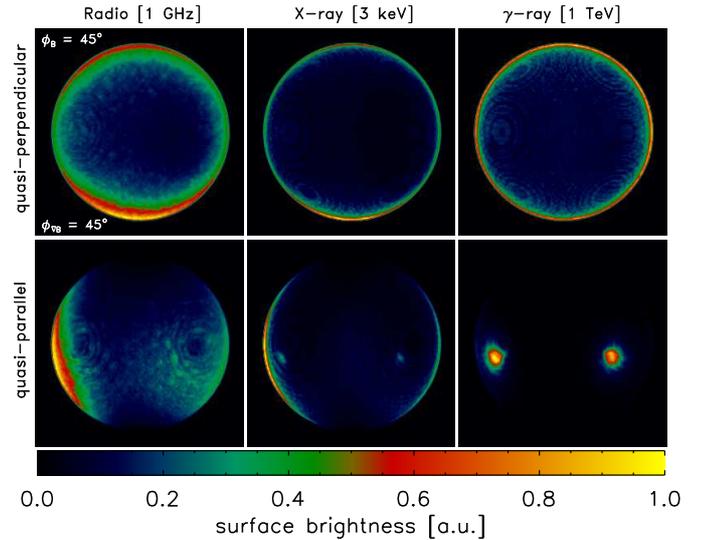}
  \caption{Maps of synchrotron radio (left), X-ray (center), and IC
           $\gamma$-ray (right) surface brightness synthesized from
           run Grad-BZ-g1, assuming quasi-perpendicular (top), and
           quasi-parallel (bottom) injection models. The adiabatic
           index is $\gamma = 4/3$. The maps have been
           synthesized adopting model B of $E\rs{max}$ shown in
           Fig.~\ref{fig8bis}. The relevant angles are $\phi\rs{B} =
           45^0$ and $\phi\rs{\nabla B} = 45^0$. The
           angle between {\avemag} and $\nabla |$\avemag$|$ is $90^0$.}
  \label{fig11}
\end{figure}

When the $\nabla B$ is not aligned with the average ambient magnetic
field (for instance in the case of run Grad-BZ-g1), the projection of the
$\nabla B$ in the plane of the sky has (for generic values of $\phi\rs{B}$
and $\phi\rs{\nabla B}$) a component perpendicular to the projected lobes
and one running between them. In this case both kind of asymmetries (lobes
converging on one side and with different brightness) are expected in the
remnant morphology. As an example, Fig.~\ref{fig11} shows the synchrotron
radio, X-ray, and IC $\gamma$-ray images synthesized from run Grad-BZ-g1,
for different injection models. The relevant angles are $\phi\rs{B} =
45^0$ and $\phi\rs{\nabla B} = 45^0$.

\subsection{Dependence on the value of $E\rs{max,\|}$}
\label{var_emaxp}

\begin{figure}[!t]
  \centering
  \includegraphics[width=8.cm]{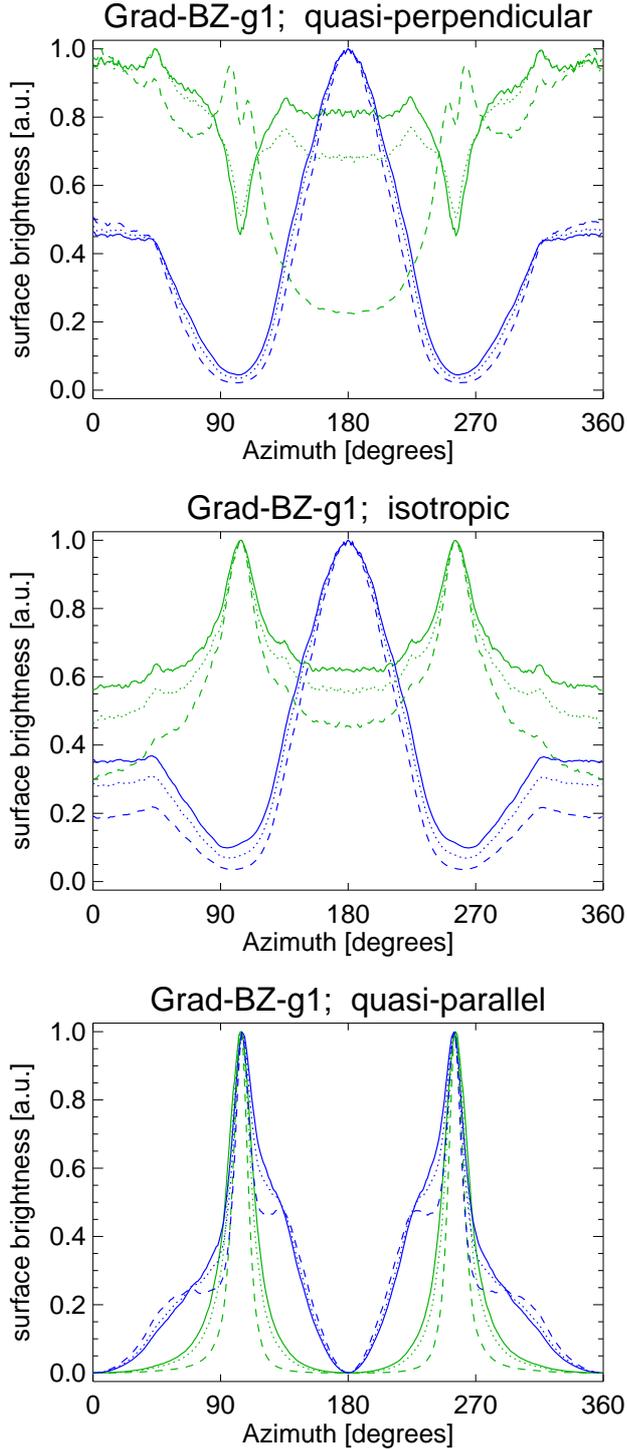}
  \caption{As in Fig.~\ref{fig6}, for the azimuthal profiles
           of the synchrotron X-ray (blue) and IC $\gamma$-ray (green)
           surface brightness synthesized from run Grad-BZ-g1, for
           $E\rs{max,\|} = 26$ TeV (solid), 5 TeV (dotted), and 1 TeV
           (dashed). The surface brightness is synthesized
           deriving the maximum energy $E\rs{max}$ as described in
           Sect.~\ref{s_unif}; its azimuthal profile is similar to that
           shown in Fig.~\ref{fig1} (solid line) but with a different
           value of $E\rs{max,\|}$. In all the cases, $E\rs{f,\|} =
           12$ TeV.}
  \label{fig12}
\end{figure}

The calculations presented above assume the free parameter
$E\rs{max,\parallel} = 26$ TeV larger than the fiducial energy
$E\rs{f,\parallel}=12$ TeV, implying that the electron energy losses
are mainly due to radiative losses (see Sect.~\ref{ps_el_distr}). The
variation of the energy distribution $N(E,a,t)$ of electrons in
Eq.~\ref{deriv_distrib} is influenced by radiative losses of electrons
that are, therefore, important for the surface brightness distribution
of the remnant in X-rays and $\gamma$-rays. Consequently, the choice
of $E\rs{max,\parallel}$ may influence both the degree and the
pattern of asymmetry of the remnant morphology. In particular, in
cases with $E\rs{max} < E\rs{f}$, we expect that $\cal{E}\rs{rad}
\rightarrow$ 1 (i.e. electrons have negligible radiative losses) and
the electron energy losses are mainly due to adiabatic expansion. This
issue is investigated by considering the reference case discussed in
Sect.~\ref{s_unif} and two additional cases for which $E\rs{max} <
E\rs{f}$, namely $E\rs{max,\parallel} = 5$ TeV and $E\rs{max,\parallel}
= 1$ TeV. Figure~\ref{fig12} shows the azimuthal profiles of the
synchrotron X-ray and IC $\gamma$-ray surface brightness synthesized
from run Grad-BZ-g1, for these values of $E\rs{max,\|}$ together with
the case with $E\rs{max,\parallel} = 26$ TeV (the reference case). Note
that $E\rs{max}$ is calculated at each point of the domain as described
in Sect.~\ref{s_unif} but for different values of $E\rs{max,\parallel}$
(see also Sect.~\ref{max_energy}). The figure shows that
for decreasing values of $E\rs{max,\|}$, the contrast of emission
increases, the effect being the largest for IC $\gamma$-ray emission
than for synchrotron X-rays. Nevertheless, the degree and the pattern
of asymmetry of the remnant morphology induced by the gradient of ISMF
are only slightly influenced by the value of $E\rs{max,\|}$.

\section{Summary and conclusions}
\label{sec5}

We developed a numerical code (\REMLIGHT) to synthesize the synchrotron
radio, X-ray, and IC $\gamma$-ray emission from MHD simulations, in the
general case of a remnant expanding through a non-uniform ISM and/or
a non-uniform ISMF. As a first application of \REMLIGHT, we coupled the
synthesis code to the MHD model discussed in Paper I (extended to
include an approximate treatment of upstream magnetic field amplification
and the effect of shock modification due to back reaction of accelerated
CRs) and investigated the effects of a non-uniform ISMF on the remnant
morphology in the X-ray and $\gamma$-ray bands. Our findings lead to
several conclusions:

\begin{itemize}

\item A gradient of ISMF strength induces asymmetries in both the X-ray
and $\gamma$-ray morphology of the remnant if the gradient has a component
perpendicular to the LoS. In general, the asymmetries are analogous to
those found in Paper I in the radio band, independently from the models of
electron injection and of maximum energy of electrons accelerated by the
shock. In the $\gamma$-ray band, the asymmetry in the remnant morphology
is inverted with respect to those in the radio and X-ray bands if
the model of $E\rs{max}$ depends inversely on the pre-shock magnetic
field strength and its contrast is ${\cal C}\rs{max} > 1$ (e.g. model B
in Fig.~\ref{fig8bis}): the brightest $\gamma$-ray lobe is located
where both the radio and the X-ray lobes are the faintest.

\item The non-thermal lobes are characterized by different brightness
when a gradient of ISMF strength is perpendicular to the lobes; they
are converging on one side when a gradient of ISMF is running between
them. In the general case of a gradient with components parallel and
perpendicular to the lobes, both kinds of asymmetry may characterize
the remnant morphology.

\item The non-thermal X-ray emission is confined in very thin limbs
because of the large radiative losses at high energy and, in general, is
the most sensitive to non-uniform ISMF. In fact the remnant morphology
in this band shows the highest degree of asymmetry among the images
synthesized in the three bands of interest (i.e. radio, X-ray, and
$\gamma$-ray), except when $E\rs{max}$ depends inversely on the
pre-shock magnetic field strength. In the latter case, the asymmetries
in the X-ray band can be significantly reduced.

\item The IC $\gamma$-ray emission is weakly sensitive to the
non-uniform ISMF, the degree of asymmetry being the lowest in the
three bands considered. The remnant morphology is almost ring-like for
quasi-perpendicular injection, shows the ``limb-inverse'' property
discussed by \cite{2009MNRAS.395.1467P} for isotropic injection
(i.e. bright $\gamma$-ray lobes correspond to dark radio and X-ray areas),
and is bilateral for quasi-parallel injection. The ``limb-inverse''
property implies, for instance, that $\gamma$-ray lobes are symmetric
and converging on one side when radio and X-ray lobes have different
brightness (see Fig.~\ref{fig4}). In case $E\rs{max}$
depends inversely on the pre-shock ambient magnetic field strength,
the asymmetries in the IC $\gamma$-ray morphology can be inverted;
for instance, brightest $\gamma$-ray lobes can be located where both radio
and X-ray lobes are fainter. Note that the $\gamma$-ray morphology of the
SNRs RX J1713.7-3946 (\citealt{2006A&A...449..223A}) and RX J0852.0-4622
(\citealt{2007ApJ...661..236A}) could be reproduced in the equatorial-belt
scenario (the injection is either quasi-perpendicular or isotropic),
whereas the morphology of SN\,1006 (\citealt{2010A&A...516A..62A}) is
compatible with that predicted in the polar-caps scenario (quasi-parallel
injection).

\end{itemize}

Note that, although the MHD model presented here does not include
self-consistently shock modification and magnetic field amplification,
we adopted an approximate treatment of both processes. Magnetic field
amplification could result from streaming instability excited by
the accelerated particles upstream of the shock or, alternatively,
the magnetic fields could be amplified in a purely hydrodynamic way in
the downstream plasma (\citealt{2007ApJ...663L..41G}). In both cases,
the shock is expected to be modified due to the dynamical reaction of the
amplified magnetic field (see, for instance, \citealt{2010A&A...509L..10F}
for an hydrodynamic model including back-reaction of accelerated CRs). In
this paper, we approach the effect of shock modification by considering
different values of the adiabatic index $\gamma$ (namely, 5/3, 4/3,
1.1) and the effect of upstream magnetic field amplification by
considering the ambient magnetic field strength enhanced by $\times 10$
in the neighborhoods of the remnant (the unperturbed field strength
commonly expected is a few $\mu$G). The main effect of $\gamma$ is
to change the compression ratio of the shock and the distance of the
contact discontinuity from the blast wave position.  In the simplest
case considered here, namely the modification on $\gamma$ and
the upstream magnetic field amplification are both isotropic with no
dependence on the obliquity angle, we found that the modified $\gamma$
and the amplified field influence mainly the absolute values of
non-thermal emission but not the large scale morphology of the remnant and
the pattern of asymmetries induced by a non-uniform ISMF. The
results presented here, therefore, are only valid in this case.
Conversely, we expect a significant effect of the modified $\gamma$ as
well as of the amplified field on the remnant morphology if the shock
modification and/or upstream magnetic field amplification depend on the
obliquity. This issue deserves further investigation in future studies.

It is worth to emphasize that the calculations provided in this paper (and
implemented in the \REMLIGHT\ code) to synthesize the non-thermal emission
from MHD simulations consider a generic adiabatic index $\gamma$. The
\REMLIGHT\ code therefore can be easily coupled with a model including the
back-reaction of accelerated CRs and synthesize the non-thermal emission
consistently if the value of an ``effective'' $\gamma$ is provided in each
point of the spatial domain\footnote{See, for instance,
\cite{2010A&A...509L..10F} for an hydrodynamic model calculating the
``effective'' $\gamma$ in each point of the spatial domain (see also
\citealt{2004A&A...413..189E})}.

Note also that the MHD model adopted here follows the evolution of the
remnant during the adiabatic phase and, therefore, its applicability is
limited to this evolutionary stage. In the radiative phase, the high
degree of compression suggested by radiative shocks leads to increase
in the synchrotron emission brightness due to compression of ambient
magnetic field and electrons. Since our model neglects the radiative
cooling of the shocked gas, it is limited to compression ratios derived
from $\gamma$ and, therefore, it is not able to simulate this mechanism
of limb brightening. Nevertheless, the model is appropriate to describe
young SNRs that are those from which non-thermal emission is commonly
detected.

\bigskip
\acknowledgements{We thank an anonymous referee for the careful reading
of the manuscript and for constructive and helpful criticism. This work
was supported in part by the Italian Ministry of University and Research
(MIUR) and by Istituto Nazionale di Astrofisica (INAF). The software used
in this work was in part developed by the DOE-supported ASC / Alliance
Center for Astrophysical Thermonuclear Flashes at the University of
Chicago. The simulations have been executed at the HPC facility (SCAN)
of the INAF-Osservatorio Astronomico di Palermo and at CINECA (Bologna,
Italy) in the framework of the INAF-CINECA agreement ``High Performance
Computing resources for Astronomy and Astrophysics".}

\bibliographystyle{aa}
\bibliography{references}

\appendix
\section{Calculation of integral in Eq.~\ref{e_rad}}
\label{app1}

The function ${\cal{I}}(a,t)$ in Eq.~\ref{e_rad} is expressed as
(\citealt{1998ApJ...493..375R})

\begin{equation}
{\cal{I}}(a,t) = \sigma\rs{B}^2 \int_{t\rs{i}}^{t}
\frac{B\rs{eff}(a,t^{\prime})^2}{B\rs{eff,s}(t)^2}
\left(\frac{\rho(a,t^{\prime})}{\rho(a,t)}\right)^{1/3}
\frac{dt^{\prime}}{t}~,
\label{integ-I}
\end{equation}

\noindent
where $B\rs{eff}^2 = B^2+B\rs{CMB}^2$ is the ``effective'' magnetic
field introduced to account for the energy losses of electrons due to
IC scatterings on the photons of CMB.

The integral (\ref{integ-I}) is rather CPU consuming because it requires
to know, with high enough time resolution, the history of each parcel of
gas inside the SNR since its shocking time. To reduce the computational
cost, we calculate it approximately, changing integration on $dt^{\prime}$
to $dR^{\prime} = V\rs{sh}(t^{\prime})dt^{\prime}$, where $R$ and
$V\rs{sh}$ are the shock position and velocity, respectively, and using
some MHD properties of the fluid.

We calculate ${\cal{I}}(a,t)$ using an analytic description of mass
density and magnetic field evolution inside the SNR which expands
through a non-uniform ISM and/or ISMF. The continuity equation
$\rho\rs{o}(a)a^2da = \rho(a)r^2dr$ results in

\begin{equation}
\rho(a,t) = \rho\rs{o}(a)\left(\frac{a}{r(a,t)}\right)^2
r\rs{a}(a,t)^{-1}
\end{equation}

\noindent
where $r\rs{a}(a,t)$ is the derivative of $r(a,t)$ with respect
to $a$; the density term in Eq. \ref{integ-I} is

\begin{equation}
\frac{\rho(a,t^{\prime})}{\rho(a,t)} =
\frac{r(a,t)^2}{r(a,t^{\prime})^2}
\frac{r\rs{a}(a,t)}{r\rs{a}(a,t^{\prime})}~.
\end{equation}

\noindent
The magnetic field in Eq. \ref{integ-I} can be expressed as $B(a,t)^2 =
B_{\parallel}(a,t)^2+B_{\perp}(a,t)^2$, where $B_{\parallel}$
and $B_{\perp}$ are the components of magnetic field parallel and
perpendicular to the shock normal, respectively. These two components
follow the magnetic flux conservation $B_{\parallel}d\sigma\rs{S} =
const$, where $d\sigma\rs{S}$ is a surface element, and the flux-frozen
condition $B_{\perp}(r)rdr = const$:

\begin{equation}
B_{\parallel}(a,t) = B_{\parallel,o}(a)\frac{a^2}{r(a,t)^2}~,
\end{equation}

\begin{equation}
B_{\perp}(a,t) = B_{\perp,o}(a)\frac{r(a,t)}{a}
\frac{\rho(a,t)}{\rho\rs{o}(a)} =
B_{\perp,o}(a)\frac{a}{r(a,t)r\rs{a}(a,t)}~.
\end{equation}

\noindent
Thus, the magnetic field and the mass density in Eq. \ref{integ-I}
can be expressed through the relation $r(a,t)$ between Eulerian
and Lagrangian coordinates of a parcel of gas and its derivative,
$r\rs{a}(a,t)$. Considering that $r(a,t)$ and $r\rs{a}(a,t)$ can
be expressed in terms of the dynamical characteristics of the shock
(i.e. as $r(a,R)$ and $r\rs{a}(a,R)$), the integral \ref{integ-I} may
be calculated as follows:

\begin{equation}
{\cal{I}}(a,t) = \frac{\sigma\rs{B}^2}{t} \int_{R\rs{i}}^{R}
\frac{B(a,R^{\prime})^2}{B\rs{s}(R)^2}
\left(\frac{r(a,R)^2 r\rs{a}(a,R)}{r(a,R^{\prime})^2
r\rs{a}(a,R^{\prime})}\right)^{1/3}
\frac{dR^{\prime}}{V\rs{sh}(R^{\prime})}~.
\label{I_approx}
\end{equation}

\noindent
Now, the relation $r(a,R)$ is approximated\footnote{The approximation
(\ref{approx}) is developed to give exact values of derivatives up to
the third order at the shock and to the first order at the center.  },
using the method described by \cite{1999A&A...344..295H}:

\begin{equation}
\frac{r(a,R)}{R} = \left(\frac{a}{R}\right)^{\psi}
(1+a\rs{1}\upsilon+a\rs{2}\upsilon^2+a\rs{3}\upsilon^3+a\rs{4}\upsilon^4)
\label{approx}
\end{equation}

\noindent
where $\upsilon = (R-a)/R$ and $\psi = (\gamma-1)/\gamma$. The parameters
$a\rs{1}$, $a\rs{2}$, $a\rs{3}$, and $a\rs{4}$ are expressed as:

\begin{equation}
a\rs{1} = -r\rs{a,s}+\psi~,
\label{a1}
\end{equation}

\begin{equation}
a\rs{2} =
\frac{1}{2}\left(Rr\rs{aa,s}-2\psi r\rs{a,s}+\psi(\psi+1)\right)~,
\end{equation}

\begin{equation}
a\rs{3} = \frac{1}{6}\left(-R^2r\rs{aaa,s}+3\psi R r\rs{aa,s} -
3\psi(\psi+1) r\rs{a,s} + \psi(\psi+1)(\psi+2)\right)~,
\label{a3}
\end{equation}

\begin{equation}
a\rs{4} = {\cal{C}} - (1+a\rs{1} + a\rs{2} + a\rs{3})~,
\end{equation}

\noindent
where ${\cal{C}}$ reflects the variation of $r(a)$ around the center of
the SNR. We adopt ${\cal{C}}={\cal{C}}\rs{A}$ where ${\cal{C}}\rs{A}$ is
given by the self-similar Sedov solution for a spherical shock (for
details see Sect.~4.3 and Appendix in \citealt{2000A&A...357..686P} and
references therein):

\begin{equation}
{\cal{C}}\rs{A}=
\left[
\frac{\gamma}{\gamma+1}\bar P(0)^{-1/\gamma}
\right]^{1/3}~,
\end{equation}

\noindent
$\bar P(0)$ is the plasma pressure at the center of the remnant divided by its 
post-shock value

\begin{equation}
\bar P(0)=
\left(\frac{1}{2}\right)^{6/5}
\left(\frac{\gamma+1}{\gamma}\right)^{6/5-\gamma/(2-\gamma)}
\left(\frac{(2\gamma+1)(\gamma+1)}{\gamma(7-\gamma)}
\right)^{(-2+5/(2-\gamma))\cdot\zeta}~,
\end{equation}

\begin{equation}
 \zeta=\frac{\gamma+1}{3(\gamma-1)+2}-\frac{2}{5}+\frac{\gamma-1}{2\gamma+1}~.
\end{equation}

\noindent
Thus, we derive ${\cal{C}}\rs{A}(\gamma = 5/3) = 1.083$,
${\cal{C}}\rs{A}(\gamma = 4/3) = 1.055$ and ${\cal{C}}\rs{A}(\gamma
= 1.1) = 1.021$. The expressions for the derivatives
$r\rs{a,s},\,r\rs{aa,s},\,r\rs{aaa,s}$ in Eqs. \ref{a1}-\ref{a3}
as functions of $R$, $\dot R$, $\ddot R$ and $R^{(3)}$ are given in
Appendix A2 of \cite{1999A&A...344..295H}.

Finally, we calculate $V\rs{sh}(R)$ in Eq. \ref{integ-I} as well as
$\ddot R$ and $R^{(3)}$, using the \cite{1987Afz....26..113H} approximate
analytical formula for the strong shock in a non-uniform medium (see
also Sect. 2.1 in \citealt{1999A&A...344..295H}).

\begin{figure}[!t]
  \centering
  \includegraphics[width=8.5cm]{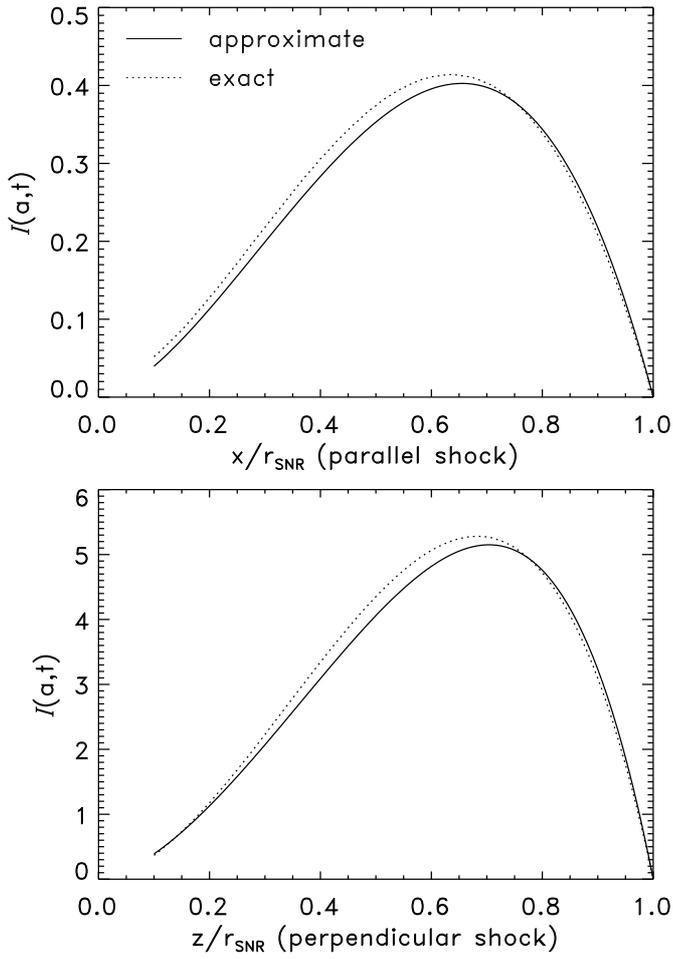}
  \caption{Self-similar approximate and exact radial profiles of the
           integral ${\cal{I}}(a)$ when the ambient magnetic field is
           either parallel (upper panel) or perpendicular (lower panel)
           to the shock normal, and $\gamma = 5/3$.}
  \label{fig_app}
\end{figure}

The integral ${\cal{I}}(a,t)$ can be calculated rather simply in the
case of a SNR expanding through uniform ISM and ISMF. We therefore
test our calculation of ${\cal{I}}(a,t)$ by comparing the approximate
values derived from Eq. \ref{I_approx} with the exact ones derived from
the Sedov solution in the case of $\gamma = 5/3$. Figure~\ref{fig_app}
compares the exact and approximate values of ${\cal{I}}(a,t)$ in the
limits of parallel and perpendicular shocks. Note that the approximate
values are very accurate at radii close to the shock front where most
of the non-thermal emission originates.

\end{document}